\documentclass[reprint,groupedaddress,longbibliography,nofootinbib,amsmath,amssymb,aps]{revtex4-1}
\usepackage{graphicx}
\usepackage{dcolumn}
\usepackage{bm}
\usepackage{soul}
\usepackage{afterpage}

\usepackage{natbib}
\usepackage{float}
\usepackage{placeins}
\usepackage{dsfont}
\usepackage[utf8]{inputenc}
\usepackage{bm}
\usepackage{amssymb}
\usepackage{amsmath}
\usepackage{amsfonts}
\usepackage{bbold}
\usepackage[normalem]{ulem}
\usepackage{graphicx}
\usepackage[usenames,dvipsnames]{xcolor} 
\usepackage{dcolumn}
\usepackage{bm}
\usepackage[pdftex,
            bookmarks,
            colorlinks,
            linkcolor=red,
            citecolor=magenta,
            menucolor=black,
            urlcolor=blue,
            plainpages=false,
            pdfpagelabels,
            hypertexnames=false]{hyperref}

\newcommand{\ket}[1]{\left| #1 \right>}
\newcommand{\bra}[1]{\left< #1 \right|}
\newcommand{\braket}[1]{ \langle{{#1}}\rangle}

\let\baraccent=\= \renewcommand{\=}[1]{\stackrel{#1}{=}}

\newcommand{\eye}{\ensuremath{\mathbb{1}}}
\newcommand{\swap}{\ensuremath{\textsf{SWAP}}}
\newcommand{\Tr}{\ensuremath{\text{Tr}}}

\begin{document}
\title{Fractal, logarithmic and volume-law entangled \\non-thermal steady states via spacetime duality
}

\author{Matteo Ippoliti}
\affiliation{Department of Physics, Stanford University, Stanford, CA 94305, USA}

\author{Tibor Rakovszky}
\affiliation{Department of Physics, Stanford University, Stanford, CA 94305, USA}

\author{Vedika Khemani}
\affiliation{Department of Physics, Stanford University, Stanford, CA 94305, USA}


\begin{abstract}

The extension of many-body quantum dynamics to the non-unitary domain has led to a series of exciting developments, including new out-of-equilibrium entanglement phases and phase transitions. 
We show how a  duality transformation between space and time on one hand, and unitarity and non-unitarity on the other, can be used to realize steady state phases of non-unitary dynamics that exhibit a rich variety of behavior in their entanglement scaling with subsystem size --- from logarithmic to extensive to \emph{fractal}. 
We show how these outcomes in non-unitary circuits (that are ``spacetime-dual" to unitary circuits)
relate to the growth of entanglement in time in the corresponding unitary circuits, and how they differ, through an exact mapping to a problem of unitary evolution with boundary decoherence, in which information gets ``radiated away'' from one edge of the system. 
In spacetime-duals of chaotic unitary circuits, this mapping allows us to analytically derive a non-thermal volume-law entangled phase with a universal logarithmic correction to the entropy, previously observed in unitary-measurement dynamics. 
Notably, we also find robust steady state phases with \emph{fractal} entanglement scaling, $S(\ell) \sim \ell^{\alpha}$ with tunable $0 < \alpha < 1$ for subsystems of size $\ell$ in one dimension. 
We present an experimental protocol for preparing these novel steady states with only a vanishing density of postselected measurements via a type of ``teleportation" between spacelike and timelike slices of quantum circuits.
\end{abstract}

\maketitle


\section{Introduction\label{sec:intro}}

Breakthrough experimental advances in building quantum simulators have opened up new regimes in the study of many-body physics, by providing direct access to the dynamics of quantum systems. This has advanced our understanding of many foundational questions ranging from the onset of chaos and thermalization, to many-body localization, to the definition of phase structure out of equilibrium~\cite{DAlessio2016, Nandkishore2015, Sondhi2020, Khemani2019}. A unifying theme in these explorations has been the study of many-body quantum entanglement across a wide variety of settings -- eigenstates of stationary Hamiltonians or of periodically driven (Floquet) evolutions; out-of-equilibrium states in driven or post-quench dynamics; and more recently, steady state ensembles of \emph{non-unitary} ``monitored'' circuits which combine unitary evolution with local non-unitary measurements (the unitary part can also be dispensed with in ``measurement only'' circuits)~\cite{Li2018, Li2019, Skinner2019, Gullans2020PRX, Choi2020, NahumSkinner2020, Lavasani2020, Ippoliti2021PRX, Hsieh2020, Lavasani2020B}. 
The exploration of non-unitary dynamics is particularly topical in the age of noisy, intermediate-scale quantum (NISQ) simulators~\cite{Preskill2018}, which naturally include non-unitary ingredients in two ways: on the one hand, by the presence of (uncontrolled) environmental noise and decoherence; on the other, by allowing controlled quantum measurements \emph{during} the dynamics (a key capability for error correction in future ``fault-tolerant'' quantum computers). 

In both unitary and non-unitary cases, both the \emph{growth of entanglement in time}, as well as its \emph{spatial scaling}, may show interesting structure, including sharp phase transitions between distinct behaviors.
A paradigmatic example is the many-body localized (MBL) phase~\cite{Basko2006, Nandkishore2015, Abanin2019}, where entanglement growth is only logarithmic in time~\cite{zpp,Bardarson_2012, Abanin_2013}; this sharply transitions to a faster algebraic (though potentially sub-ballistic~\cite{Agarwal2015,Vosk2015,Luitz_2016,LezamaSlow,NahumHuse_Griffith}) growth at an MBL-to-thermalizing transition.
Monitored non-unitary circuits, on the other hand, feature sharp transitions in the spatial scaling of their steady-state entanglement~\cite{Li2018, Skinner2019, Gullans2020PRX, Choi2020, Li2019}: from an area-law (at high measurement rate) to a volume-law (at low enough rate), through a logarithmically entangled critical point described by a conformal field theory~\cite{Jian2020, Li2020, Bao2020, Zabalo2020}. In all, we are only beginning to explore the rich variety of novel phenomena displayed by the \emph{dynamics} of many-body systems and our understanding of most questions in this domain, particularly in the non-unitary setting, is still nascent. 

In this work, we add to this understanding by studying entanglement dynamics in a new class of non-unitary evolutions -- those that are ``spacetime duals'' of unitary circuits -- recently introduced by two of us~\cite{Ippoliti2021PRL}. Spacetime duality, in simple terms, exchanges the roles of space and time in a quantum evolution, and (generically) associates to every unitary circuit a non-unitary partner. Dualities reveal connections between seemingly distinct problems, and they often point to entirely \emph{new} results or phenomena---Wegner's gauge-spin duality being a paradigmatic example~\cite{Wegner}; this case is no exception.

Our work makes three main advances: 
\begin{itemize}
    \item First, we present new classes of steady state phases for entanglement dynamics. These include, most notably, a robust family of \emph{fractally entangled} steady-state phases that lie outside the established classifications of ``area-law'' / ``volume-law'' / logarithmic entanglement scaling generically exhibited by the eigenstates or steady-states of unitary quantum dynamics.\footnote{We are aware of only three special cases---critical points between non-interacting Anderson localized phases~\cite{Kopp2006, Jia2008, PotterNandkishore2014}, free fermions with fractal Fermi surfaces induced by long-range hopping~\cite{Trombettoni2015}, and Motzkin chains~\cite{MovassaghShor2016,Sugino2018}---in which eigenstates can display fractional (but not fractal) entanglement scaling. Two of these are free, and all are fine-tuned points in parameter space rather than robust phases of matter. Fractal entanglement in a different sense---that of a fractal-shaped entanglement profile---has been found in the context of Floquet non-Hermitian CFTs~\cite{Ageev2021}.}
    
    \item Second, our results are obtained via a method that is interesting in and of itself. The spacetime duality transformation allows us to build on the existing body of knowledge on entanglement dynamics in unitary circuits, and thus affords a powerful analytical handle on the study of non-unitary dynamics that was otherwise missing away from special limits\footnote{For example, the limit of large local Hilbert space dimension~\cite{Bao2020, Jian2020} or of all-to-all interactions~\cite{NahumRoy2020}.}. 
    In particular, we analytically derive a universal subleading logarithmic correction to the entanglement of non-thermal volume-law steady states that exactly realizes the conjectured universal behavior of unitary-projective circuits in the entangling phase~\cite{Fan2020, LiFisher2020}; this log-correction is key to our understanding of the volume law phase as a type of dynamically generated error correcting code, which hides information form local measurements and thus allows entanglement to survive~\cite{Choi2020, Gullans2020PRX}.
    
    \item Third, on the experimental side, our work opens up new ways for more efficient realizations of measurement-induced phases in experiments. In particular, we show how the steady states of spacetime dual circuits can be obtained with an exponentially smaller ``post-selection overhead'' (explained below), which is a central practical challenge associated with preparing many-body states that realize measurement-induced phases. 
    
    \end{itemize}

We now give a slightly more detailed outline of these ideas.
One of the main objects of this work is to explore
the relationship between entanglement in spacetime dual partner circuits---or, more precisely, between \emph{entanglement dynamics} in the unitary circuit and \emph{spatial scaling of entanglement} in late-time states of its non-unitary dual.
Unitary dynamics gives rise to a variety of possible behaviors for entanglement growth. The non-unitary dual circuits inherit this variety, and thus display a similar wealth of steady state phases characterized by different spatial entanglement scalings. Notably, some of the phases thus obtained go beyond the possibilities of generic unitary dynamics. 
Correspondingly, phase transitions in the growth of entanglement in unitary circuits (e.g. across an MBL transition) map onto phase transitions in the steady states of the spacetime-dual models. 

\begin{figure}
\centering
	\includegraphics[width=1.\columnwidth]{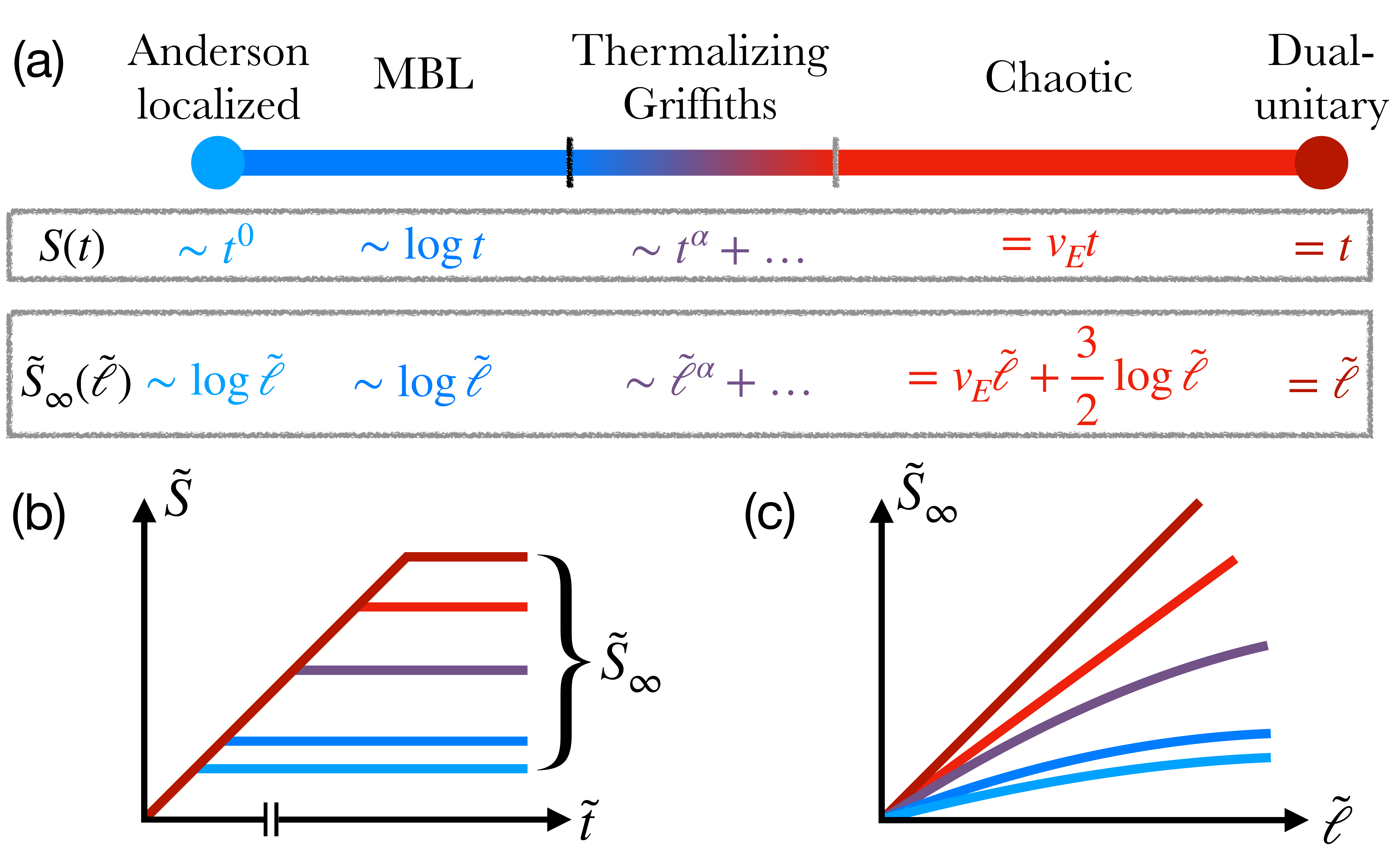}
	\caption{Summary of various dynamical regimes in unitary circuits, and properties of steady states in their spacetime duals. 
	(a) Regimes from least to most entangling.
	First row: growth of entanglement in time, $S(t)$, in unitary circuits of each type. 
	Second row: scaling of entanglement with subsystem size $\tilde{\ell}$ in the steady state of the spacetime-dual circuit, $\tilde{S}_\infty$. 
	The two agree up to logarithmic terms, which dominate for Anderson localized circuits. In the Griffiths regime, only the leading order term is shown; for other cases, the results are expected to be exact up to terms at most constant in $\tilde{\ell}$. 
	(b) Growth of $\tilde{S}$ in the spacetime-dual circuit: all regimes have ballistic growth with maximal speed at first ($\tilde{S}=\tilde{t}$), but saturate to different values. 
	(c) Saturation value $\tilde{S}_\infty$ as a function of subsystem size $\tilde{\ell}$, going from logarithmic to ballistic, with arbitrary power-law ($\tilde{S}_\infty \sim \tilde{\ell}^\alpha, \, 0<\alpha<1$) in between. 
	}
	\label{fig:phases}
\end{figure}

A schematic summary of our work is shown in Fig.~\ref{fig:phases}.
We sweep through several classes of unitary evolutions, displaying the full range of behaviors for entanglement growth: from Anderson localized circuits (where entanglement saturates to a constant~\cite{Bardarson_2012}), to Floquet MBL~\cite{Abanin_2015,Abanin_2016,Zhang2016} circuits (where it grows logarithmically~\cite{zpp, Bardarson_2012,Huse_2014,Abanin_2013}), then past an MBL-to-thermal phase transition to a slowly thermalizing phase (where entanglement is thought to grow sub-ballistically, as $\sim t^\alpha$ with $0<\alpha<1$, due to rare disorder-induced bottlenecks, so called ``Griffiths'' effects~\cite{Agarwal2015,Vosk2015,Luitz_2016,LezamaSlow, NahumHuse_Griffith}), all the way to generic chaotic circuits (where entanglement grows ballistically~\cite{KimHuse_2013,HoAbanin_entanglement,Mezei2017}). 
To leading approximation, each type of entanglement growth is mirrored in a distinct phase of a non-unitary circuit, characterized by  the spatial scaling of entanglement at late times which ranges from logarithmic to fractal to volume-law. 

Most strikingly, the sub-ballistic entanglement growth in disordered, thermalizing systems translates to a \emph{fractal} scaling of entanglement in steady states of a robust, generic class of non-unitary circuits. In these states, entropy scales as a fractional, tunable power law of subsystem size (in one dimension), and is statistically self-similar over all length scales.
These fractally entangled steady states are generically not obtained as either eigenstates or dynamical steady states in any unitary setting, and represent a striking example of new, robust nonequilibrium phenomena made possible by adding non-unitarity to the toolkit of many-body quantum dynamics.

Separately, spacetime duality, as a theoretical tool, affords us a great degree of analytic tractability, which is key in understanding all these types of steady-state entanglement scaling.  
Specifically, we show that for a unitary circuit $U$, the entanglement properties of states evolved by its (non-unitary) spacetime-dual circuit $\tilde{U}$ can be related to entanglement induced by $U$, with the roles of space and time exchanged, but with an additional twist: the unitary evolution is accompanied by \emph{edge decoherence}, which allows information to escape the system and be ``radiated away" from one of its edges. This is summarized by Eq.~\eqref{eq:boundary_decoh} below.

To leading order, as mentioned earlier, the growth of entanglement in time under $U$ is mapped to the spatial scaling of steady state entanglement under $\tilde{U}$, leading to the range of different scalings summarized in Fig.~\ref{fig:phases}.
Importantly, however, the interchanging of space and time is not the full story.
The added twist of edge decoherence leaves its footprint in logarithmic contributions to the steady-state entanglement in the spacetime-dual circuit. 
This allows us to furnish an analytic derivation of universal subleading logarithmic corrections to non-thermal volume-law steady states. Identical corrections have been observed in unitary-measurement circuits~\cite{Li2019}, and found to be universal features of their non-thermal volume-law phase related to its quantum error correcting properties~\cite{Fan2020, LiFisher2020}. 
Furthermore, the logarithmic correction may become the \emph{leading} contribution in cases where one may na\"ively expect area-law steady states (e.g. in duals of Anderson-localized models). Area-law steady states are in fact ruled out in spacetime duals of unitary circuits, barring trivial exceptions. Relatedly, the mapping to edge decoherence also allows us to prove that at short (dual) ``times'', the dual circuit $\tilde{U}$ produces \emph{ballistic} growth of entanglement, at the maximum possible speed, even when the associated unitary circuit $U$ is in a localized phase.

Finally, we note that spacetime duality is not only an interesting theoretical construct; it is also experimentally motivated. A crucial experimental challenge with realizing measurement induced phases is the ``post-selection overhead'' associated with the measurements: any nontrivial entanglement phase structure is only displayed by \emph{individual} quantum trajectories~\cite{dalibard1992wave} corresponding to pure states labeled by a fixed sequences of measurement outcomes, while it is lost in a stochastic mixture over measurement outcomes~\cite{Li2018}. Experimentally measuring any observable feature of the output state requires reproducing the same state multiple times (a single experimental run corresponds to a single ``shot"); this requires us to fix (i.e. post-select) the sequence of measurement outcomes in order to reproduce a given output state, which creates a huge overhead for any finite density of measurements (exponential in the spacetime volume of the circuit)\footnote{In some cases, it may by possible to overcome this barrier with the aid of classical computation, by solving a suitable ``decoding'' problem (for efficiently simulable evolutions, e.g. non-interacting or Clifford circuits, as well as for local probe qubits coupled to general circuits, where one can levarage locality~\cite{Gullans2020PRL, Noel2021}); 
however, we expect the decoding problem for the full many-body wavefunction to be hard in the worst case. The postselection overhead then offers a useful upper bound to the practical cost of realizing these states.}.
As was shown in Ref.~[\onlinecite{Ippoliti2021PRL}], the postselection cost can be significantly ameliorated (or, sometimes, eliminated altogether) when computing the purity of a (mixed) density matrix evolving under the spacetime-duals of unitary circuits.
A generalization of this idea also applies to the present context, where the (pure) steady states of interest can be prepared by using only local unitary gates and a limited number of postselected measurements scaling as the \emph{boundary} of the circuit in spacetime---an exponential improvement over more conventional unitary-measurement setups~\cite{Li2018, Skinner2019}. 
We achieve this by ``teleporting'' the input and output states of dual circuits, which live on (experimentally unnatural) timelike surfaces, to spacelike surfaces that can be more readily accessed in experiments. We also discuss a family of experimentally realizable disordered Floquet Ising models with an MBL phase transition, whose spacetime duals display the full gamut of steady state entanglement phases discussed above. 

The balance of the paper is organized as follows. In Section~\ref{sec:setup}, we review the notion of spacetime duality and discuss how to realize spacetime duals of unitary circuits experimentally using only a vanishing density of projective measurements in spacetime. 
We then introduce the mapping to edge decoherence that serves as the main theoretical tool for the rest of the work. 
In Section~\ref{sec:anderson} we use this mapping to discuss duals of (Anderson- and many-body) localized circuits and argue that steady-state entanglement diverges logarithmically with subsystem size in both cases. 
In Section~\ref{sec:Haar} we turn to generic chaotic evolution, modeled by Haar-random circuits; we obtain volume law entangled steady states, but with a universal non-thermal logarithmic correction stemming from the edge decoherence. 
Finally, in Section~\ref{sec:fractal} we consider slowly thermalizing models with sub-ballistic entanglement growth, mapped by spacetime duality to fractally entangled steady states. 
We summarize our results and discuss their implications for future research in Section~\ref{sec:conclusion}.


\section{Setup\label{sec:setup}}

\subsection{Non-unitary dynamics from spacetime duality}

\begin{figure}
\centering
	\includegraphics[width=\columnwidth]{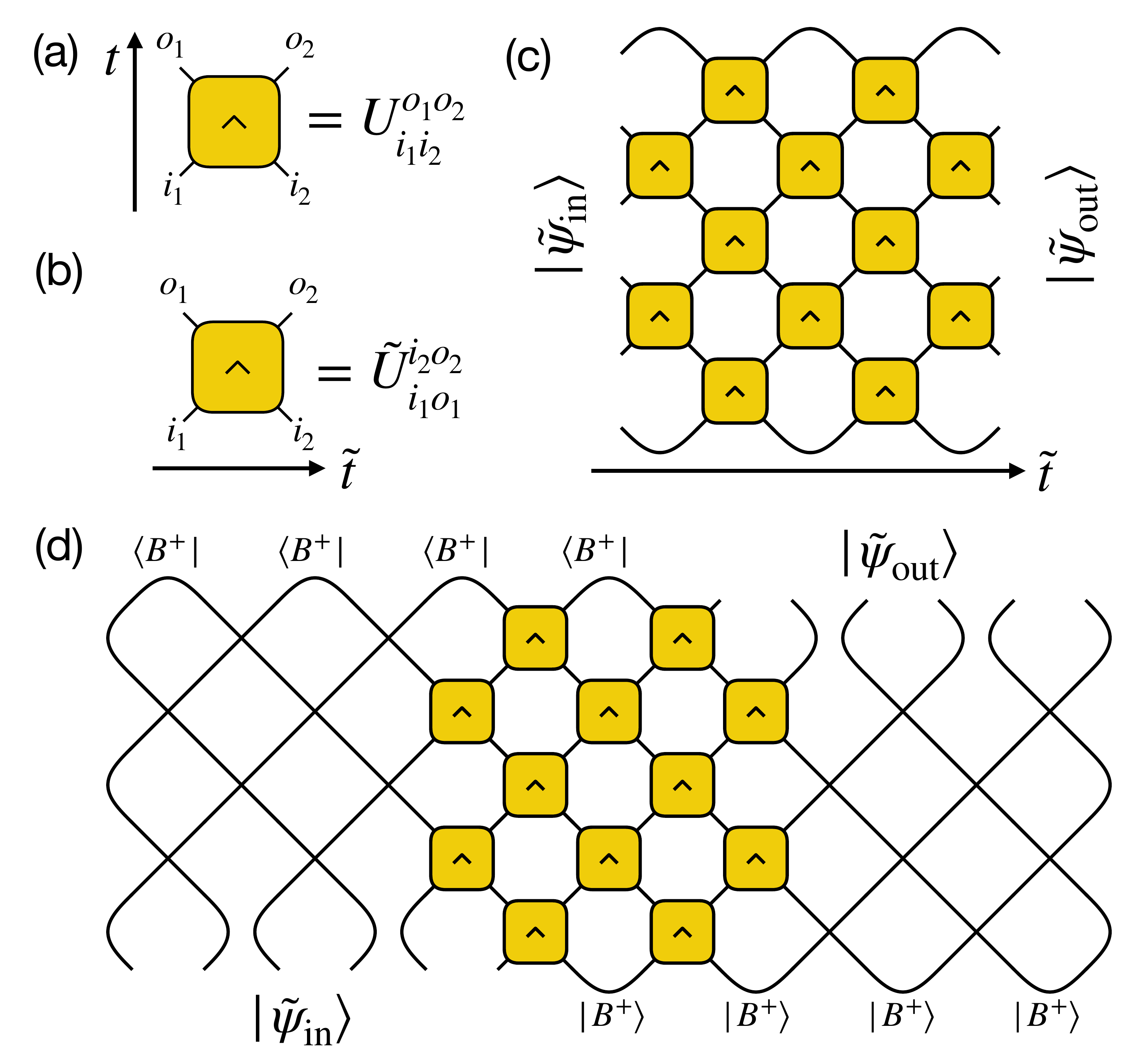}
	\caption{Spacetime duality. 
	(a) A two-qudit unitary gate $U$. Bottom legs are inputs, top legs are outputs. The caret symbol on the gate denotes the direction of unitarity.
	(b) If we follow the arrow of time $\tilde{t}$ (left to right), the same object describes a non-unitary operation $\tilde{U}$. 
	(c) Spacetime dual of a brickwork unitary circuit: the non-unitary gates $\tilde{U}$ turn the input state $| \tilde{\psi}_\text{in}\rangle$ (at the left boundary) into the output state $| \tilde{\psi}_\text{out}\rangle $ (at the right boundary). 
	Both states exist on timelike surfaces. 
	(d) The input and output states can be ``teleported'' to spacelike surfaces by employing ancillas initialized in a Bell pair state $\ket{B^+} = q^{-1/2}\sum_{i=1}^q \ket{ii}$, $\swap$ gates, and postselected projective measurements on $\bra{B^+}$.
	The number of postselected measurements is proportional to the length of the circuit's boundary.
	}
	\label{fig:flipping}
\end{figure}

To review the idea of spacetime duality~\cite{Ippoliti2021PRL}, we start from the simplest instance -- that of a two-qudit unitary gate. Fig.~\ref{fig:flipping}(a,b) illustrates how this one object, $U_{i_1i_2}^{o_1o_2}$ (unitarily mapping two inputs $i_{1,2}$ to two outputs $o_{1,2}$ according to ``arrow of time'' $t$), could alternatively be viewed ``sideways'', according to a rotated ``arrow of time'' $\tilde{t}$, as mapping input qubits $i_1,o_1$ to output qubits $i_2,o_2$.
The resulting map, $\tilde{U}_{i_1o_1}^{i_2o_2}$---which we will call the ``(spacetime)-dual'' or ``(spacetime)-flipped'' version of $U$---is in general \emph{not} unitary.

For example, if $U$ is a two-site identity gate, its dual $\tilde{\eye} = q\ket{B^+}\bra{B^+}$ is proportional to a projection onto the Bell pair state $\ket{B^+} \equiv q^{-1/2} \sum_{a=1}^{q} \ket{aa}$ ($q$ is the local Hilbert space dimension). 
Notice that this is a \emph{forced} measurement: the measurement outcome is fixed.
In general, a polar decomposition yields $\tilde{U} = q F W$, where $W$ is a unitary gate and $F$ is positive semi-definite and normalized to $\Tr(F^2) = 1$. 
Since $F\geq 0$, we can interpret $\tilde{U}$ as an element of a POVM: it corresponds to a \emph{forced weak measurement} (i.e., deterministically postselecting a particular outcome of a POVM); alternatively, one could think of $F$ as finite-time imaginary time-evolution with a two-site Hamiltonian.

We note that the use of forced measurements (rather than Born-random quantum measurements) does not obstruct the existence of measurement-induced entanglement phases and transitions.
Deep in an entangling phase, with very infrequent measurements, outcomes are expected to be uniformly distributed, so that forced measurements should be equivalent to typical trajectories of random measurements~\cite{Fan2020}.
In Clifford circuits, measurement outcomes yield ``phase bits'' that do not affect entanglement~\cite{Aaronson2004, LiChenFisher_2018} (provided the outcomes are mutually compatible\footnote{Forcing incompatible measurement outcomes results in the annihilation of the state.}).
More generally, forced and random measurements are thought to give rise to the same qualitative phenomenology (with specific quantitative differences in criticality~\cite{NahumRoy2020} and late-time dynamics~\cite{Fidkowski2020}).

Having defined how spacetime duality acts on individual gates, it is straightforward to extend the idea to $1+1$ dimensional circuits made out of two-site unitary gates in a ``brick-wall'' pattern~\cite{Ippoliti2021PRL}. 
This idea of flipping circuits is at the core of much recent work on a variety of topics. In particular, important analytical progress on quantum chaos has been achieved via special \emph{dual-unitary} circuits~\cite{Akila_2016,Prosen_SFF1,Prosen_SFF2,Prosen_entanglement,Gopalakrishnan_2019,Prosen_correlations,Prosen_OpEnt,Piroli_2020, Claeys2020, Prosen_correlations2}, where each gate $U$ making up the circuit is such that $\tilde{U}$ happens to be unitary as well.
More generically, other works have explored the idea of using the spacetime-flipped circuits as a tool for calculating properties of the original unitary evolution, with applications ranging from tensor network contractions~\cite{Banuls_2009,Hastings2014,Harrow2020,Potter2020} to MBL~\cite{Chalker_MBL,Abanin_MBL,Chan2020} to chaos and thermalization~\cite{Chan2018,Chalker_Feynman,Abanin_influence}. On the other hand, Ref.~[\onlinecite{Ippoliti2021PRL}] (by a subset of us) introduced the idea of using spacetime duality to a different end---not to understand features of the associated unitary evolution, but rather to study the non-unitary dynamics in its own right. Here we build on this to engineer new phases and phenomena, such as exotic non-thermal steady states.

For consistency, when drawing circuit diagrams we will always choose the direction of (unitary) time $t$ as bottom to top (also indicated by a small arrowhead symbol on each gate), while the ``dual'' arrow of time $\tilde{t}$ flows left to right\footnote{A tilde is used to designate quantities in the flipped circuit.}.
The spacetime dual of a unitary circuit thus evolves states ``sideways'', left to right.

This may seem to pose a conceptual issue, since the input and output states, $|\tilde{\psi}_\text{in/out}\rangle$ (both pure), exist on \emph{timelike} (vertical) slices of the circuit, Fig.~\ref{fig:flipping}(c).
However this can be remedied, as shown in Fig.~\ref{fig:flipping}(d). The idea is to ``teleport" the input and output states from their native timelike surface to spacelike ones~\footnote{The term is inspired by quantum teleportation, where Alice and Bob share a Bell pair and Bob can perform some local operations (including measurements) to ``teleport'' a desired state to Alice's qubit. Here too we initialize an array of Bell pairs, feed one qudit from each pair into a unitary circuit, and then at the final time measure one part of the system to produce the desired output on the other part.} by using ancillary qudits arranged in 1D (the half of these on the right are initialized in Bell pair states $\ket{B^+}$), a brick-wall pattern of $\swap$ gates that extends the original unitary circuit on the left and right\footnote{A similar procedure of ``swapping in'' additional qubits at the edges was considered in Ref.~[\onlinecite{Li2020}].}, and measurements only at the end of the unitary time-evolution. The upshot is that an experimentalist can obtain the desired target output state of the dual non-unitary evolution by simply performing a (suitably modified) unitary evolution, followed by a set of measurements only at the final time. 
These final measurements implement the ``upside-down'' Bell pairs at the top of the circuit in Fig.~\ref{fig:flipping}(d), which represent open boundary conditions for the flipped circuit\footnote{Another way of stating this is to note that open boundary conditions are equivalent to having a special bond where all gates are $\eye$; using $\tilde{\eye} \propto \ket{B^+}\bra{B^+}$ yields the result. A subtlety with the wavefunction normalization is discussed in Appendix~\ref{app:edge_d}.}.

As mentioned above, these Bell measurements are forced, or postselected -- the experimentalist performs a Bell measurement, and discards any realizations where the outcome is not $|B^+\rangle$. 
However, we note two advantageous features of this setup, compared to more generic unitary-measurement circuits. 
First, the measurements all take place at the end of the circuit, rather than in the middle; this is useful as some quantum simulator architectures do not allow measurements in the middle of a circuit (measurements can be deferred to the end of the circuit, at the expense of injecting a growing number of ancillas\cite{Noel2021}). 
Second, the number of forced/postselected measurements scales only with the circuit's \emph{boundary}, rather than its spacetime volume.
Thus spacetime duality presents a way to realize interesting non-unitary circuits with a drastically reduced postselection overhead. 

This overhead is further ameliorated by a suitable choice of initial states: if $|\tilde{\psi}_\text{in}\rangle$ is set to be a product of nearest-neighbor Bell pairs, then we can get rid of the ancillas on the left in Fig.~\ref{fig:flipping}(d), since this particular choice of initial condition is automatically realized by simply taking open boundary conditions on the left edge of the associated unitary circuit, as depicted in Fig.~\ref{fig:boundary_dissip}(a). 
In the following, we will focus on such Bell-pair initial states. In addition to lowering the postselection overhead, this ensures that, in the thermodynamic limit $\tilde{L}\to\infty$, the state $|\tilde{\psi}_\text{out}\rangle$ maintains its normalization at all times, despite the flipped circuit being non-unitary, as we prove in Appendix~\ref{app:edge_d}. 

\subsection{Mapping to boundary decoherence}\label{sec:boundary_decoh}

We now discuss the connection between entanglement dynamics in the unitary circuit and spatial scaling of entanglement in its spacetime-dual. This connection is made crisp by an exact mapping to a problem of unitary evolution with ``edge decoherence'' that we present below, with some further technical details in Appendix~\ref{app:edge_d}.

\begin{figure}
\centering
	\includegraphics[width=1.\columnwidth]{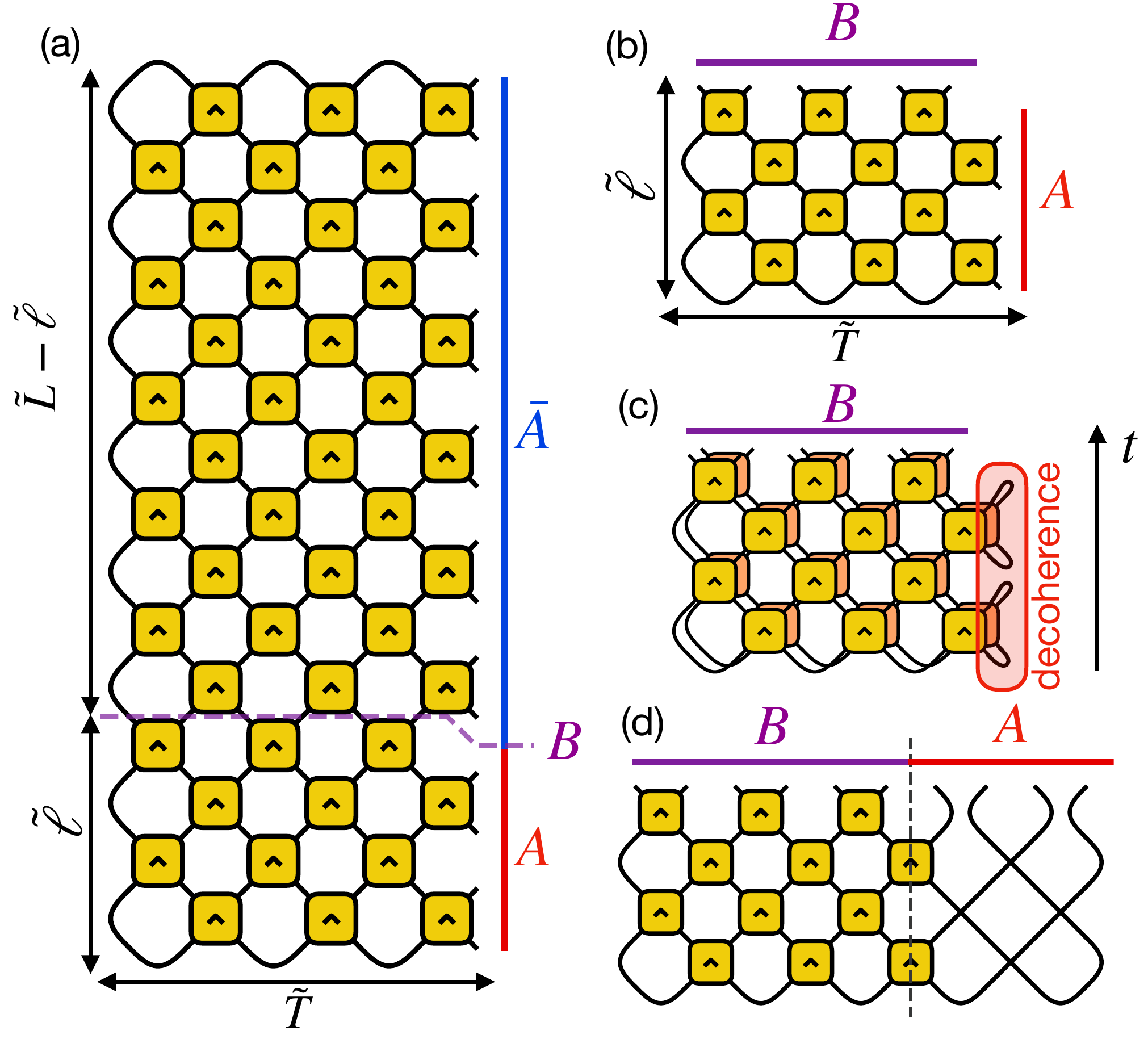}
	\caption{Mapping the entropy of the spacetime-dual circuit's output state to that of unitary dynamics with boundary decoherence. 
	(a) Setup: we fix the input state $|\tilde{\psi}_\text{in}\rangle$ to be a product of Bell pairs (open boundary, left), the output state $|\tilde{\psi}_\text{out}\rangle$ lives on the the right boundary. We consider an entanglement cut between a subsystem $A$ (of size $\tilde{\ell}$) and its complement $\bar{A}$ (of size $\tilde{L}-\tilde{\ell}$). The depth of the non-unitary circuit is $\tilde{T}$.
	We also highlight a spacelike surface $B$ along at the entanglement cut $t=\tilde{\ell}$.
	(b) In the limit $\tilde{L}\to\infty$, with $\tilde{\ell}$ finite, system $B$ is isometrically encoded in system $\bar{A}$: for the purpose of computing entropy, the part of the circuit above cut $B$ can be elided.
	(c) Reduced density matrix on subsystem $B$. The tracing out of $A$ can be interpreted as the action of decoherence (via a fully depolarizing channel). $\rho_B$ is the output of $\tilde{\ell}$ layers of unitary dynamics \emph{and} decoherence on the edge qudit.
	(d) Equivalently, one can ``teleport'' subsystem $A$ from a timelike surface to a spacelike one by using $\tilde{\ell}$ ancillas (initialized in Bell pair states) and $\swap$ gates, and compute the entropy of the resulting pure state about the cut between $A$ and $B$ (dashed line).}
	\label{fig:boundary_dissip}
\end{figure}

Before we begin, let us fix some conventions and notation. We will denote by $L$ and $T$ the size (number of qudits) and depth (number of gate layers) of the unitary circuit, and by $\tilde{L}$ and $\tilde{T}$ the size and depth of its (non-unitary) spacetime dual; these obey $L=\tilde{T}$ and $T = \tilde{L}$.
We will denote by $\tilde{S}$ the entanglement entropy of a timelike (vertical) subsystem (for instance, region $A$ in Fig.~\ref{fig:boundary_dissip}(a)) and by $S$ the entropy of a spacelike (horizontal) subsystem (for instance region $B$ in Fig.~\ref{fig:boundary_dissip}(a)).
The on-site Hilbert space dimension is $q$ (in most cases below, $q=2$). 

Our main goal is to understand the entanglement properties of the late-time output states of the flipped circuit. 
We will focus on the entropy of contiguous subsystems near one edge, denoted by $A$, with size $|A| \equiv \tilde{\ell}$. The complementary subsystem, $\bar{A}$, consists of $\tilde{L}-\tilde{\ell}$ sites.
This setup is illustrated by Fig.~\ref{fig:boundary_dissip}(a). We will denote the (von Neumann or R\'enyi) entropy of region $A$ by $\tilde{S}(\tilde{T},\tilde{\ell})$.

The setup is considerably simplified if we take the thermodynamic limit $\tilde{L}\to\infty$. As we show next, in this limit the entropy of $A$ can be given a very useful alternative characterization in terms of the original unitary dynamics coupled to a bath that induces decoherence at one of its edges. 
The main steps in deriving this mapping go as follows (additional technical details are given in Appendix~\ref{app:edge_d}):
\begin{enumerate}
    \item Consider the set of legs denoted by $B$ in Fig.~\ref{fig:boundary_dissip}(a), which live on a spacelike surface at (unitary) time $t=\tilde{\ell}$. As we show in Appendix~\ref{app:edge_d}, when $\tilde{L} \to \infty$, the information in $B$ is isometrically encoded in subsystem $\bar{A}$ (the thermodynamically large complement of $A$) on the right boundary\footnote{This argument relies on the fact that we fixed $|\tilde{\psi}_\text{in}\rangle$ to be a product of Bell pairs. Otherwise, information could also ``leak out'' at the left boundary.}. This means that the information shared between $A$ and $\bar{A}$ is the same as the information shared between $A$ and $B$; we can thus get rid of the entire part of the tensor network above $B$, reducing the problem to a much smaller circuit of dimensions $\tilde{\ell} \times \tilde{T}$ shown in Fig.~\ref{fig:boundary_dissip}(b).
    \item Because the state on $AB$ is pure, the entropy of $A$ can be obtained by tracing out either subsystem; we trace out $A$ and obtain the reduced density matrix of $B$, $\rho_B$. This amounts to taking two copies of the circuit (a ``bra'' and a ``ket'') and connecting them along $A$. Viewed in the time direction $t$, after each layer of unitary gates the rightmost site is traced out and replaced by a maximally mixed state. This corresponds to a fully depolarizing channel~\cite{NielsenChuangBook} acting on the right edge of the system, as illustrated in Fig.~\ref{fig:boundary_dissip}(c). $\rho_B$ is a mixed state which is the output of this combined unitary-decoherence evolution.
    \item An equivalent picture is obtained by  ``teleporting'' the right edge of the circuit to a spacelike slice by using ancillas and $\swap$ gates as discussed earlier. 
    As shown in Fig.~\ref{fig:boundary_dissip}(d), this gives an enlarged unitary circuit, split in two subsystems $A$ and $B$ where the evolution looks very different (a $\swap$ circuit on $A$, the original unitary gates on $B$).
\end{enumerate}

The upshot is that the spectrum of $\rho_A$ is the same as that of a state $\rho_B$ obtained by evolving the system with $t=\tilde{\ell}$ layers of unitary gates, intercalated by \emph{fully depolarizing noise} at an edge qubit\footnote{This is true up to some zero eigenvalues, which arise due to the fact that $A$ and $B$ have different sizes and do not contribute to any of the R\'enyi entropies.}. In formulae:
\begin{equation}\label{eq:boundary_decoh}
    \lim_{\tilde{L}\to\infty} \tilde{S}(\tilde{T},\tilde{\ell}) = S_\text{dec}(t=\tilde{\ell},L=\tilde{T}),
\end{equation}
where we used $S_\text{dec}$ to denote the entropy of the mixed state evolving with edge decoherence. 

First, let us consider the early-time dynamics of the flipped circuit, when $\tilde{T} \ll \tilde{\ell}$. This corresponds to running the boundary-depolarized dynamics for a time that is very long compared to the size of the system ($t=\tilde\ell\gg \tilde{T} = L$). 
As we argue in Appendix~\ref{app:edge_d}, unless the unitary gates are highly fine-tuned, this dynamics has a unique steady state, which is completely mixed~\footnote{Logarithms are taken base $q$.}: $\lim_{t\to\infty} S_\text{dec}(t,L) = L$. By Eq.~\eqref{eq:boundary_decoh}, this means that at early times, the entropy of $\rho_A$ grows \emph{ballistically}, at the maximal entanglement velocity allowed by the brick-wall geometry: $\tilde{v}_E = 1$.

The rest of the paper will be devoted to understanding what value the entropy saturates to at late times: $\tilde{S}_\infty(\tilde{\ell}) \equiv \lim_{\tilde{T}\to\infty} \lim_{\tilde{L}\to\infty} \tilde{S}(\tilde{T},\tilde{\ell})$ (in fact, $\tilde{T} > \tilde{\ell}$ is sufficient for saturation). 
By Eq.~\eqref{eq:boundary_decoh}, this is equal to $S_\text{dec}(t=\tilde{\ell})$ in the thermodynamic limit $L\to\infty$.
The spatial scaling of late-time entanglement in the flipped circuit is thus mapped to the propagation of decoherence from the edge into the bulk of a unitarily-evolving system, as a function of time.
This is the main result of this section. 
As it turns out, this is indeed closely related to (but subtly different from) the growth of half-chain entropy in a closed system evolving under the original unitary circuit.

We now provide some intuition on the behavior of $S_\text{dec}(t)$ and argue that, at least for chaotic models, it is similar to the growth of entanglement in a \emph{purely unitary} circuit, without edge decoherence. 
To see why, it is useful to consider the evolution of the reduced density matrix $\rho_{B}$ from an operator spreading perspective~\cite{HoAbanin_entanglement,Mezei2017}. 
In this formulation, one considers expanding the density matrix of the full state in a basis of Pauli strings; the reduced density matrix of $B$ is simply given by those strings that are supported entirely within $B$. 
As operators spread out under unitary dynamics, strings that were initially contained in $B$ will eventually develop support outside it, increasing the state's entropy.
We can now compare two situations: (i) an infinite unitary circuit, with the same type of gates applied everywhere and (ii) the one depicted in Fig.~\ref{fig:boundary_dissip}(d), where the gates outside of $B$ have been replaced with swap gates. Within $B$, the two are clearly the same. The difference is that in case (ii), once a Pauli string's endpoint leaves $B$, it is ``radiated'' away by the swap gates, with no chance of ever returning\footnote{In other words, those Pauli strings are killed by the edge decoherence. This also shows why the maximally mixed steady state should be unique: other steady states would only be possible if there were operators that \emph{never} reach the edge, e.g., if the unitary dynamics had \emph{exactly} conserved quantities with no support at the edge. This occurs only in highly fine-tuned cases.}. This is in contrast with case (i), where operators have the possibility to shrink and re-enter $B$. 
Since such shrinking processes are rare -- at least for sufficiently chaotic dynamics -- we can expect (i) and (ii) to behave similarly. 

To summarize: we showed that, given a unitary circuit $U$, the entropy of a contiguous region of size $\tilde{\ell}$ in the late-time output state of the (non-unitary) dual circuit $\tilde{U}$ is the same as that of a semi-infinite chain evolving under $U$ \emph{and} boundary decoherence for time $t=\tilde{\ell}$. 
This latter description is in turn related to the growth of half-chain entanglement entropy under unitary dynamics, except that operators that straddle the cut are never allowed to shrink. Since the shrinking of operators is already rare under chaotic dynamics, one might expect this difference to be negligible. 
We will show below that while this is true to leading order for chaotic dynamics, an important difference nevertheless appears in the form of subleading logarithmic contributions. These can, in turn, become the leading contribution for localized dynamics, where the unitary evolution produces entanglement very slowly, or not at all.

Before moving on to applications of this mapping, we note that the above derivation applies to the case when the subsystem $A$ is at the edge of a half-infinite chain.
Alternatively, we could consider a situation where the chain is infinite on both sides, and $A$ corresponds to a block of sites in its bulk. In that case, the reduction of the circuit---from Fig.~\ref{fig:boundary_dissip}(a) to (b)---can be done on both sides, both above and below $A$. 
The resulting circuit has a less transparent interpretation\footnote{It corresponds to a the entropy of a state $\rho_{BC}$ on a bipartite system $BC$ where $C$ serves as a ``reference system'' for $B$: the two are initially in a pure, maximally entangled state, then $B$ evolves under unitary dynamics with edge decoherence while $C$ has no dynamics.}. Nevertheless, intuitively one expects that the leading order scaling with $\tilde{\ell}$ should be the same for both physical situations. We confirm this for the Haar random circuit case in Sec.~\ref{sec:Haar}. However, we also find that the coefficient of the aforementioned logarithmic correction is \emph{different} in the two cases, a fact which we explain in terms of properties of random walks.

\subsection{Models of unitary dynamics \label{sec:models}}

To ground our discussion, it is helpful to consider a specific set of experimentally realizable unitary circuit models that, given appropriate parameter choices, can exhibit all the relevant types of entanglement dynamics.
For concreteness we consider a family of ``kicked Ising models'': one-dimensional spin chains evolving under the Floquet unitary
\begin{equation}
U_F = e^{-i \sum_n g_n X_n} e^{-i\sum_n h_n Z_n + J_n Z_nZ_{n+1}} \;,
\label{eq:kicked_ising}
\end{equation}
where the transverse fields $g$, longitudinal fields $h$, and Ising couplings $J$ may take any (clean or disordered) values. 
These models have already been at the center of many important developments in quantum dynamics, from out-of-equilibrium phases~\cite{Khemani2016, Khemani2019} to quantum chaos~\cite{KimHuse_2014,Prosen_SFF1,Prosen_SFF2}.
While Eq.~\eqref{eq:kicked_ising}, as written, is in the form of a time-dependent Hamiltonian, it can be easily recast into a brickwork circuit of two-qubit gates (i.e. it can be ``Trotterized'' exactly). These models can be realized in Rydberg atoms or digital simulators such as Google's Sycamore processor~\cite{Google2019, Ippoliti2020}. 
Spacetime duality then allows us to associate a non-unitary circuit to each such unitary circuit.

The model in Eq.~\eqref{eq:kicked_ising} can realize the entire range of entanglement growth regimes summarized in Fig.~\ref{fig:phases}, sorted here from slowest to fastest: 
\begin{itemize}
\item[(i)] Floquet-Anderson localization, where $S(t)$ saturates to an area-law in in $O(1)$ time. Achieved by setting $h\equiv 0$ (which makes the model free) and having disorder in $J$, $g$.
\item[(ii)] Many-body localization (MBL), with slow logarithmic entanglement growth, $S(t)\sim \log(t)$. A parameter manifold where the MBL phase in this model has been studied~\cite{Zhang2016} is $g \equiv 0.72\Gamma$, $J \equiv 0.8$, $h_n \equiv 0.65 + 0.72\sqrt{1-\Gamma^2}G_n$, where $G_n$ are standard normal variables. The model has an MBL-to-thermal transition at $\Gamma_c\simeq 0.3$.
\item[(iii)] Strongly-disordered thermalizing dynamics near the MBL transition, where entanglement growth is \emph{sub-ballistic}, $S(t)\sim t^\alpha$~\cite{LezamaSlow}. Realized e.g. by the above model at $\Gamma\gtrsim 0.3$.
\item[(iv)] Dynamics deep in the ergodic phase, where entanglement growth is ballistic, $S(t)\sim v_E t$, with finite ``entanglement velocity'' $v_E$~\cite{KimHuse_2013,HoAbanin_entanglement,Mezei2017,Nahum2017,Nahum2018,Keyserlingk2018}. Realized e.g. by the above model at $\Gamma\simeq 1$. 
\item[(v)] Dual-unitary dynamics. Realized by setting $g = J = \pi/4$, for arbitrary $h_n$. This is a provably chaotic model (sometimes called ``maximally chaotic") with maximum entanglement velocity~\cite{Prosen_SFF2}. 
\end{itemize}

We will begin by examining the extremes of this range--localized dynamics in Sec.~\ref{sec:anderson} and generic chaotic dynamics (including the dual-unitary case) in Sec.~\ref{sec:Haar}.
We will then analyze the sub-ballistic regime in Sec.~\ref{sec:fractal}. In the latter two cases, rather than trying to simulate the kicked Ising chain directly (which suffers from finite size limitations), our analysis will focus on coarse-grained models with analytic and numerical tractability that are expected to reproduce the same qualitative behavior in the appropriate regimes.


\section{Logarithmically entangled steady states\label{sec:anderson}}

The discussion in Sec.~\ref{sec:setup} highlights a close but subtle relationship between (i) \emph{entanglement growth in unitary circuits} and (ii) \emph{scaling of late-time entanglement} in their spacetime-duals. 
We begin our exploration of this relationship in a class of models where the difference is sharpest: free-fermion Floquet-Anderson localized circuits where the unitary evolution is periodic in time, non-interacting and disordered. 

In Floquet-Anderson localized circuits, eigenmodes of the Floquet unitary $\hat{b}_n$ are given by superpositions of on-site fermionic modes $\hat{a}_n$ with exponentially-decaying envelopes: $\hat{b}_m \equiv \sum_n \psi^{(m)}_{n} \hat{a}_n$,  $|\psi^{(m)}_{n}| \sim e^{-|x_m-n|/\xi}$ ($\xi$ is the single-particle localization length\footnote{In principle, this could differ between modes; here we take $\xi$ to be some characteristic typical localization length.}, 
$x_m$ is the position of the $m$-th orbital's center).
Since there are no interactions (and thus no dephasing), entanglement growth about a cut originates entirely from orbitals straddling the cut; this leads to saturation to an area-law at late times: $S(t)\sim t^0$ for $t\gg 1$.
One may expect this behavior to translate to area-law entangled steady states in the spacetime-dual circuit.
However, as we show next, entanglement in these models saturates instead to a \emph{logarithmically divergent} value.
This provides an example of a more general result -- that entanglement \emph{cannot} saturate to an area-law in spacetime-duals of unitary circuits, except for some trivial fine-tuned exceptions.
This is a straightforward consequence of the aforementioned fact that $\rho_B$ (the output of unitary dynamics with edge decoherence, sketched in Fig.~\ref{fig:boundary_dissip}(c)) eventually has to reach an infinite temperature state, as we prove in Appendix~\ref{app:edge_d}.

\begin{figure}
    \centering
    \includegraphics[width=\columnwidth]{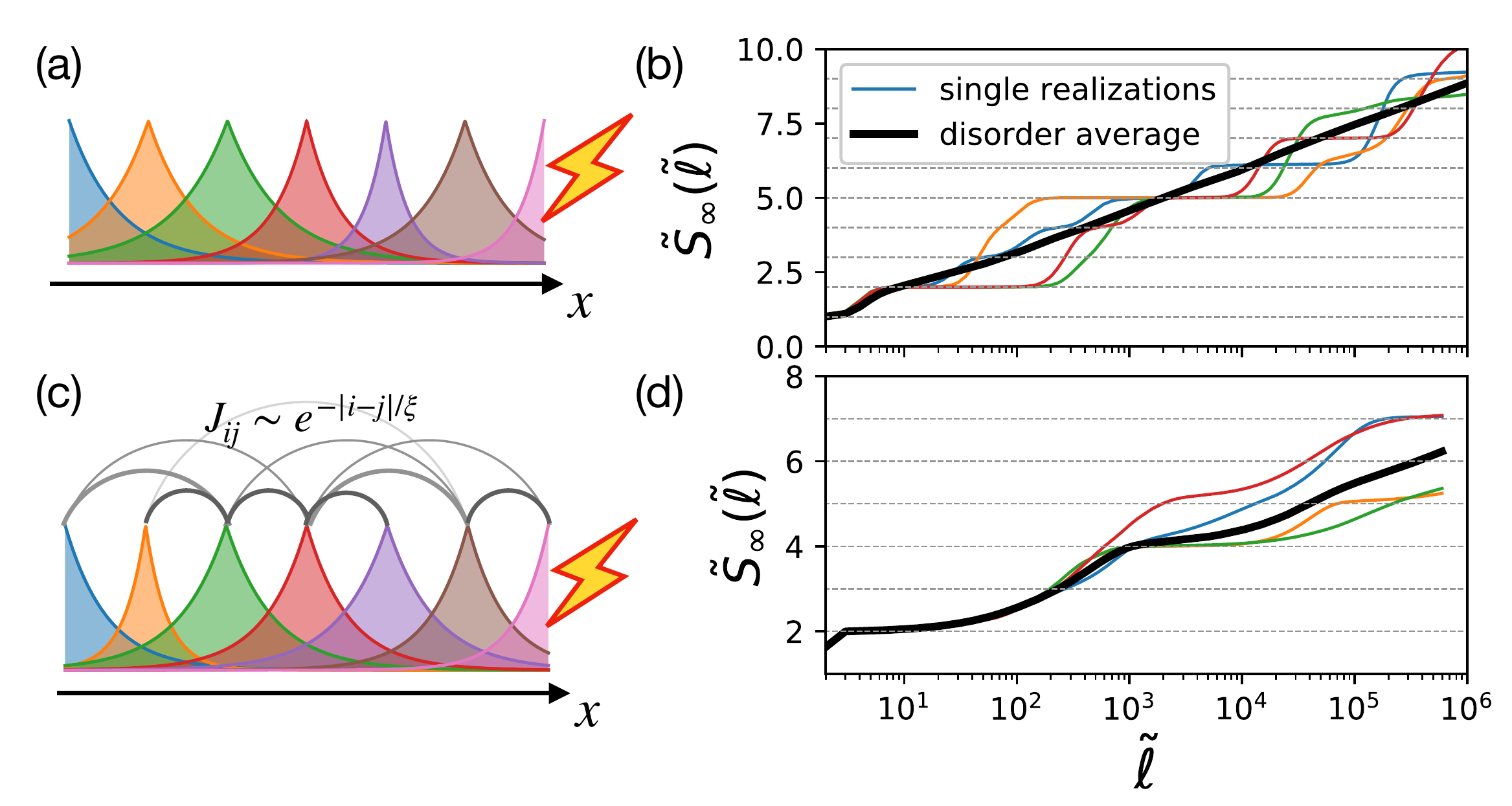}
    \caption{Logarithmic entanglement in spacetime-duals of localized circuits. 
    (a) Schematic of Floquet-Anderson orbitals with with edge decoherence. Each exponentially localized orbital decoheres over a timescale $\tau_n$ dictated by its overlap with the edge site. 
    (b) Results of numerical fermionic Gaussian state simulations (see details in main text).
    The disorder-averaged entropy $\tilde S_\infty$ grows logarithmically with $\tilde\ell$, while individual realizations show steps associated to the decoherence of individual orbitals. 
    (c) Similar sketch for Floquet MBL problem; the difference with (a) is that the exponentially localized l-bits $\tau^z_i$ interact with each other.
    (d) Entropy vs subsystem size $\tilde{S}(\tilde\ell)$ for the dual of the Floquet MBL model at $\Gamma = 0.1$ (see text). Full density matrix evolution of $\tilde{T}=9$ qubits evolving under the Floquet MBL circuit with edge decoherence. The behavior is qualitatively similar to that of the Floquet-Anderson problem, with steps and plateaus associated to the depolarization of individual l-bits.
    \label{fig:anderson}}
\end{figure}

We can obtain the logarithmic scaling of $\tilde{S}_\infty(\tilde{\ell})$ from the mapping in Eq.~\eqref{eq:boundary_decoh}.
We therefore consider the density matrix $\rho_B$ obtained by evolving a pure initial state with $t\equiv \tilde{\ell}$ layers of the unitary Floquet-Anderson circuit \emph{and} a depolarizing channel on one of the edge sites (as sketched in Fig.~\ref{fig:anderson}(a)). 
The overall process is a fermionic Gaussian map and can be studied within the free-fermion formalism -- one can, in particular, compute its (complex) eigenvalues and eigenmodes, and from them compute the entanglement $S_\text{dec}(t)$ (note that the initial Bell pair state is itself Gaussian). Numerical results are shown in Fig.~\ref{fig:anderson}(b) for the Ising model~\eqref{eq:kicked_ising} in the non-interacting limit $h_n =0$ ($J_n\equiv \pi/3$ and $g_n$ maximally disordered in $[0,2\pi]$); they clearly show a logarithmic growth of the average entropy, with single realizations additionally showing an interesting step-like structure. 

The result can be understood intuitively as follows. Imagine starting with the purely unitary circuit, with its localized eigenmodes, and continuously switching on the depolarizing noise at the boundary. The orbitals of the unitary model are affected by the noise only to the extent that they overlap with the noisy site, which is an exponentially small effect for orbitals localized far from the boundary. Therefore, the $m$-th orbital picks up a small imaginary contribution to its energy, of size $\sim e^{-|x_m|/\xi}$, which sets a time scale $\tau_m \sim e^{|x_m|/\xi}$ for the orbital to be decohered by the boundary. At time $t$, all orbitals with $\tau_m\ll t$ have fully decohered, and each of them contributes one bit of entropy to $\rho(t)$;
conversely, orbitals with $\tau_m\gg t$ remain unaffected by the decoherence and contribute no entropy. As a result, we have $S(t)\sim \xi \log(t)$.
We conclude that late-time states in the spacetime-dual circuit have entanglement scaling $\tilde{S}_\infty(\tilde{\ell}) \sim \log{\tilde{\ell}}$, with a non-universal coefficient proportional to the localization length $\xi$ in the associated unitary circuit -- which in turn can be tuned by e.g. adjusting the disorder strength. 

A similar analysis carries over to Floquet many-body localized (MBL) systems~\cite{Abanin_2015,Abanin_2016,Zhang2016, Lazarides_2015}, with the local integrals of motion (\emph{l-bits})~\cite{Huse_2014,Abanin_lbits} playing the role of Floquet-Anderson orbitals. In particular, it is still true that the l-bits' exponential tails expose them to the effects of edge decoherence; however, in addition, the l-bits can also directly dephase each other due to their (exponentially weak) interactions, potentially accelerating the spread of entanglement in the system (see Fig.~\ref{fig:anderson}(c)). In the unitary circuit, exponentially decaying interactions between the lbits lead to a characteristic logarithmic growth of entanglement in time~\cite{zpp, Bardarson_2012,Abanin_2013}. When considering the spacetime-dual circuit, there are therefore two distinct sources of logarithmic dependence on $\tilde{\ell}$: one from na\"ively exchanging space and time, and one from the boundary decoherence that caused logarithmic growth even in the Anderson localized circuits discussed above. 
The combination of these two still results in $\tilde{S}_\infty(\tilde{\ell}) \sim \log{\tilde{\ell}}$, as confirmed numerically by the data shown in Fig.~\ref{fig:anderson}(d). 
Here, we used the version of the Ising model \eqref{eq:kicked_ising} from Ref.~[\onlinecite{Zhang2016}] (also introduced in Sec.~\ref{sec:models}), where $\Gamma$ is used to tune across an MBL-to-thermal transition (at $\Gamma_c\simeq 0.3$); we used $\Gamma=0.1$, deep in the MBL phase. 

We note that while the steady states discussed in this section exhibit a logarithmic scaling of entanglement, they otherwise do not appear to be ``critical'' (e.g. the coefficient of the logarithm is fully non-universal) and we do not expect them to exhibit conformal symmetry, unlike other recent examples of logarithmic entanglement in non-unitary systems~\cite{Saleur2017,Chen2020}. Possible connections between these states and other logarithmically-entangled, non-critical states in unitary systems (e.g. at infinite-randomness fixed points~\cite{Refael2004, Bonesteel2007}) are an interesting question for future work.


\section{Non-thermal volume-law entangled steady states \label{sec:Haar}}

Next, we turn to chaotic unitary dynamics.
Such dynamics have been fruitfully modeled by random circuits where each two-site gate is an independently chosen Haar-random unitary. 
As minimal models for chaotic quantum evolution, these circuits have been studied extensively and are known to capture various universal features of operator spreading and entanglement growth~\cite{ChandranLaumann_2015,Nahum2017,Nahum2018,Keyserlingk2018,Khemani_2018,Rakovszky_2018,Rakovszky_2019}. 
Here, we make use of these results to uncover universal features of steady-states of spacetime duals of chaotic unitary evolutions, such as the clean limit of the kicked Ising chain~\eqref{eq:kicked_ising}.

The quantity that lends itself to a particularly simple calculation in the Haar-random circuit is the so-called \emph{annealed average} of the second R\'enyi entropy, which is obtained by averaging the purity of a subsystem over realizations of the random circuit, and then taking a logarithm\footnote{Throughout this section, logarithms are taken base $q$.}: $q^{-S_2^{(a)}(t)} \equiv \overline{\mathcal{P}_A(t)}$ where the purity of the reduced density matrix $\rho_A$ is defined as $\mathcal{P}_A = \text{Tr}(\rho_A^2)$ and the overline denotes averages over circuit realizations. It is known that this quantity can be evaluated in terms of a classical random walk for the endpoint of a domain wall-like object which we define below~\cite{Nahum2018}. In our case, the boundary depolarization changes this calculation by inducing a partially absorbing boundary condition on this random walk. As a result, the purity picks up an additional factor proportional to $\sim t^{-1/2}$ (the survival probability of the random walk). Upon taking the logarithm, this gives the result
\begin{equation}\label{eq:HaarResult}
    \tilde{S}^{(a)}_{2,\infty}(\tilde{\ell}) = S^{(a)}_{2,\text{dec}}(t = \tilde{\ell}) = v_E \tilde{\ell} + \frac{1}{2} \log{\tilde{\ell}} + \ldots,
\end{equation}
where $\ldots$ stands for corrections that are at most constant in $\tilde{\ell}$. Note that the entropy density exactly coincides with the entanglement velocity in the unitary circuit, $v_E$. Moreover, there is a sub-leading logarithmic correction whose $1/2$ prefactor is fixed by the diffusive dynamics of the entanglement domain wall.

\begin{figure}
\centering
	\includegraphics[width=\columnwidth]{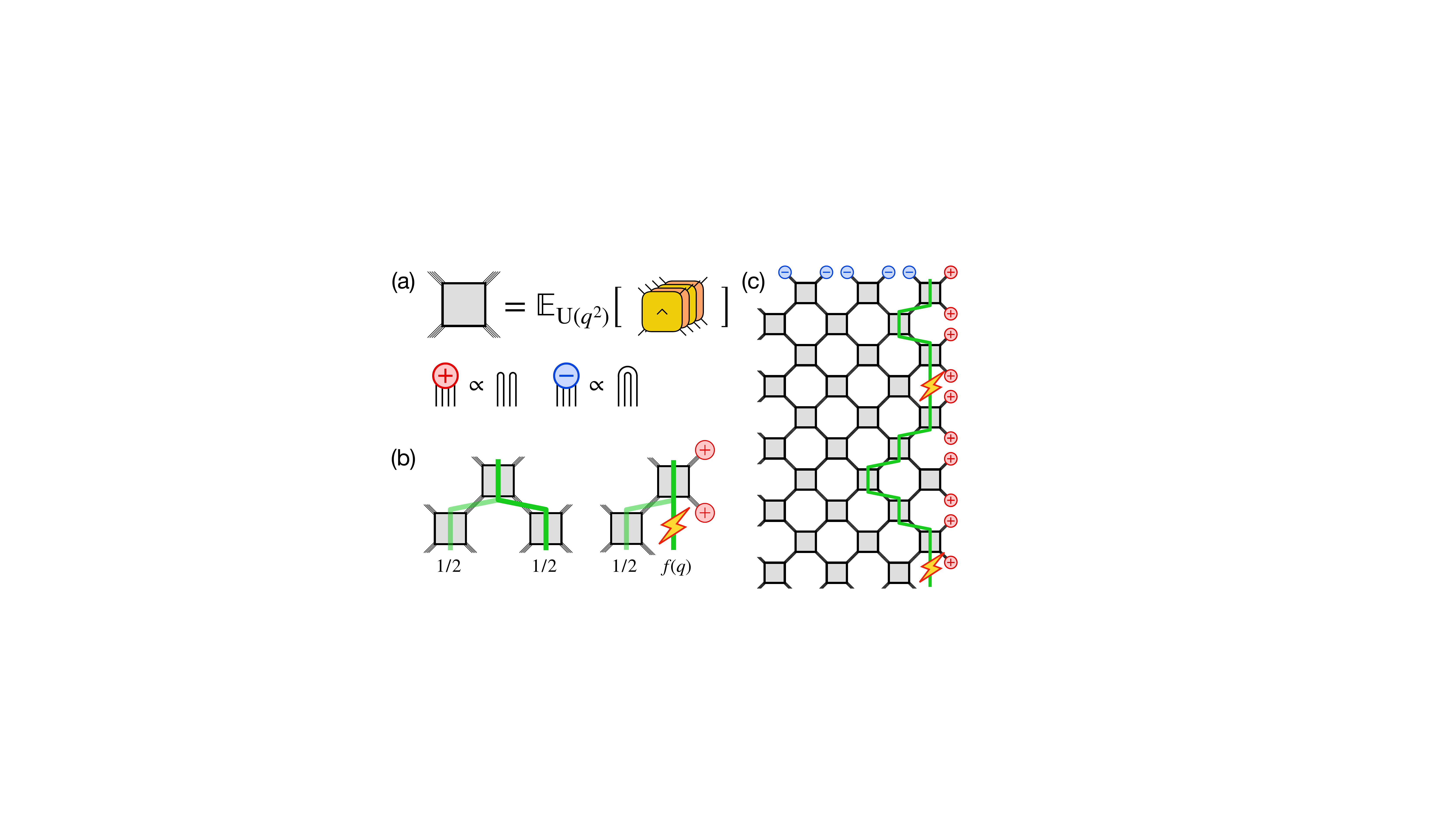}
	\caption{Random walk calculation of average purity in Haar random circuit with boundary decoherence. 
	(a) Notations: the gray boxes represent Haar-averaged unitary gates; the $\pm$ symbols represent the two ways of pairing up the four legs; the purity calculation involves a domain wall between $-$ and $+$. 
	(b) Elementary steps of the random walk in the bulk (left) and boundary (right), where probability is not conserved ($f(q)<1/2$).
	(c) A realization of the random walk. Whenever the random walker hits the boundary, its survival probability decreases (lightning symbols).
	}
	\label{fig:Haar}
\end{figure}

In order to derive Eq.~\eqref{eq:HaarResult}, let us first briefly review how the calculation of $\overline{\mathcal{P}_A(t)}$ proceeds in the original Haar random unitary circuit. Since $\mathcal{P}_A$ is quadratic in the density matrix, this calculation involves \emph{four} copes of the circuit: two ``ket'' and two ``bra'' variables for each leg. 
To get a non-zero average, one needs to pair ket and bra variables, which can be done in two inequivalent ways, denoted by $+$ and $-$ in Fig.~\ref{fig:Haar}(a). Averaging each gate over the Haar measure results in a two-dimensional tensor network that can be thought of as an Ising-like classical partition function in terms of the $\pm$ variables. On the lower boundary of this tensor network, the boundary conditions are fixed by the initial state $\ket{\psi_0}$, while on the upper boundary one should have a domain of $-$ states inside $A$ surrounded by $+$ states in its complement.

Let us focus on the case when the subsystem $A$ is half of an infinite chain. In this case, the boundary condition takes the form of a domain wall in the Ising formulation. One can then consider evaluating the full 2D tensor network row-by-row, starting from this boundary state. The great simplification arising from Haar-averaging is that at each step, the domain wall evolves into similar domain-wall configurations; when a gate is applied at the position of the domain wall, its endpoint moves randomly left or right with equal probabilities, while picking up an overall prefactor. This corresponds to a recursion relation of the purities which takes the form 
\begin{equation}\label{eq:HaarRecursion}
    \mathcal{P}(x,t) = \frac{2q}{q^2+1} \frac{\mathcal{P}(x-1,t-1) + \mathcal{P}(x+1,t-1)}{2},
\end{equation}
whenever a gate is applied on the bond between sites $x,x+1$. Here, we used $\mathcal{P}(x,t)$ to denote the purity of the subsystem $A = [-\infty,\ldots , x]$ at time $t$.
Following this process all the way back to time 0, we end up with the result
\begin{equation}\label{eq:HaarPurity}
    \mathcal{P}(x,t) = \left(\frac{2q}{q^2+1}\right)^{t} \sum_y K_{xy}(t) \mathcal{P}(y,0),
\end{equation}
where $K_{xy}(t)$ is the propagator (from position $x$ to $y$) of a simple random walk. For a product initial state\footnote{The reader might worry that we in our case, we are dealing with initial states of the Bell-pair type, rather than product states. Note, however, that at any given time $t$, the random walker can only end up either on the even or the odd sub-lattice. As such, $\mathcal{P}(y,0)$ is still the same for all $y$ appearing in Eq.~\eqref{eq:HaarPurity}.}, we have $\mathcal{P}(y,0) = 1$ for all $y$. By normalization, we also have $\sum_y K_{xy}(t) = 1$ at all times. Therefore, we end up with a simple result: $\mathcal{P}(x,t) = q^{-v_\text{E}(q)t}$, where $v_\text{E}(q) = \log((q+q^{-1})/2)$ is the \emph{entanglement velocity}\footnote{Note that this is in particular the entanglement velocity associated with the annealed average of the 2nd R\'enyi entropy -- one could call it the \emph{purity velocity}~\cite{Nahum2018,Lamacraft2018}. The rate of growth of the \emph{quenched average}, $S_2^{(q)} \equiv \overline{S_2(t)}$ is in general different, although the difference is small~\cite{ZhouNahum_2019}. The speeds associated to other entropies (e.g. von Neumann), on the other hand, can be significantly different~\cite{Keyserlingk2018}.}. 

How does this result change when we take the spacetime dual of the Haar random circuit, or, equivalently, when we add depolarizing noise at the right boundary? Inside subsystem $B$, the domain wall performs the same random walk as before. However, when it hits the boundary of $B$, the depolarizing channel imposes \emph{partially absorbing}\footnote{The fact that the boundary is only partially, not fully, absorbing comes about because the two states that make up the left/right half of the domain wall are non-orthogonal: their overlap is the factor $1/q$ picked up at the boundary.} (also known as \emph{radiation}) boundary conditions on it. Whenever the domain wall would leave $B$, it instead remains on the same position but picks up an extra factor of $1/q$. Whenever this happens, the domain wall will be ``out of sync'' with the circuit and will miss the following layer of gates. This results in an additional factor of $q^{v_E}$. Overall, this means that whenever the domain wall would hit the boundary, it picks up a factor of $2f(q) \equiv q^{v_E-1} < 1$, decreasing its total survival probability. This is illustrated in Fig.~\ref{fig:Haar}(b,c). 

\begin{figure}
\centering
	\includegraphics[width=\columnwidth]{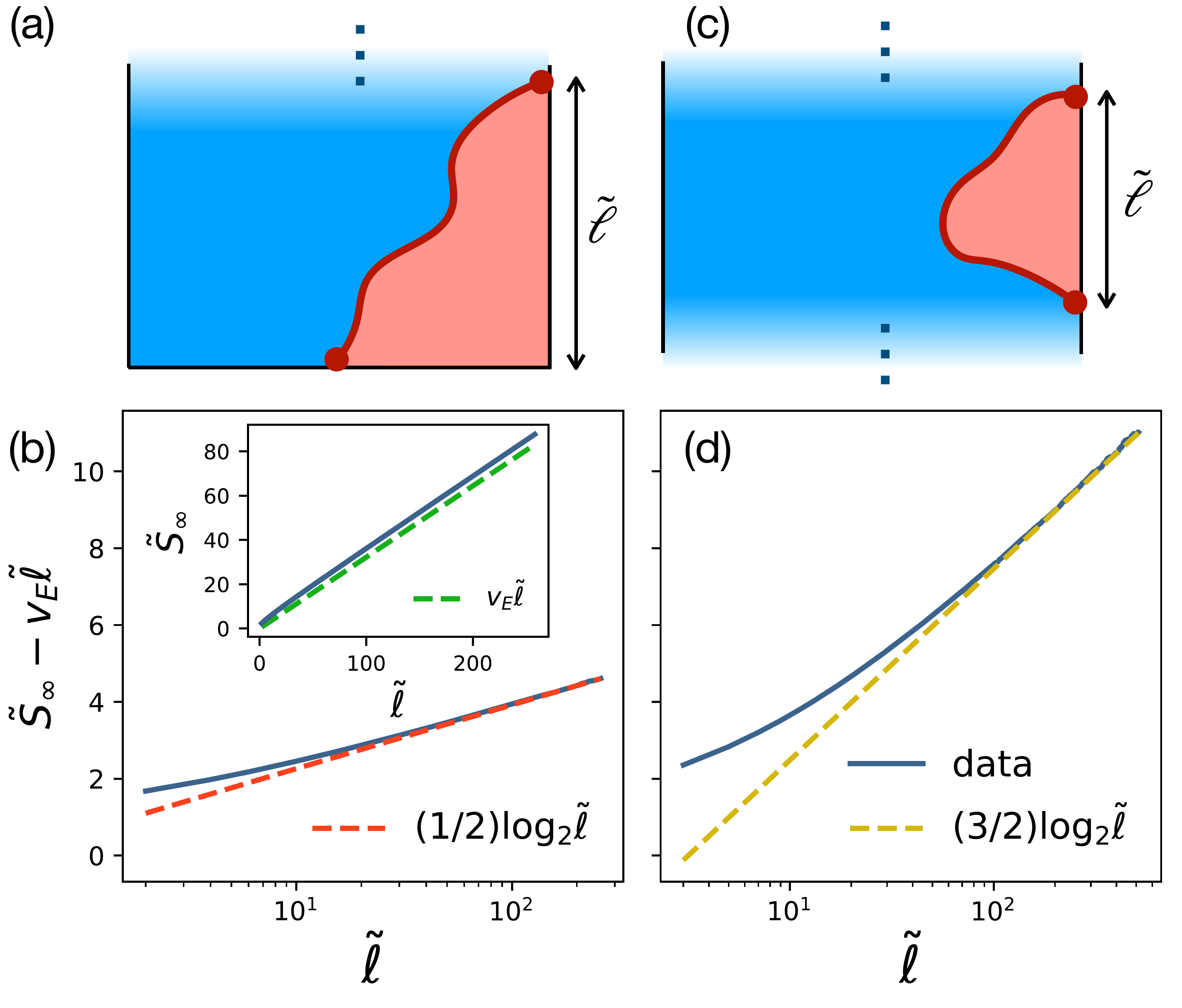}
	\caption{Entanglement entropy in spacetime duals of Haar-random circuits.
	(a) Schematic of a finite subsystem at the edge of a semi-infinite system: the entropy calculation maps to random walk survival probability (starting point is pinned to the entanglement cut, endpoint is free).
	(b) Numerical results for the entropy confirm the analytical prediction: a volume-law term $v_E\tilde\ell$ (inset; $v_E = \log_2(5/4)$) with a $(1/2)\log\tilde\ell$ non-thermal correction.
	(c) Schematic of a finite subsystem in the bulk of an infinite system: entropy calculation maps to random walk return probability (starting and ending points are pinned to the edges of the subsystem).
	(d) Numerical results agree with the predicted $(3/2)\log\tilde\ell$ correction.
    Numerical data in (b,d) is from stabilizer simulations of Clifford unitary circuits with edge decoherence with $\tilde{T} = 129$ qubits; the annealed average is over $10^6$ circuit realizations. For (d), the initial mixed state (bottom of (c)) is implemented by adding $\tilde{T}$ reference qubits with no dynamics.
	}
	\label{fig:Haar_num}
\end{figure}

Applying this logic, we find that the purity of the evolution with boundary depolarization becomes
\begin{equation}\label{eq:Purity_dissip}
    \tilde{\mathcal{P}}(x,t) = q^{-v_E(q) t} \sum_y \tilde{K}_{xy}(t) \tilde{\mathcal{P}}(y,0),
\end{equation}
where $\tilde{K}_{xy}(t)$ is now the propagator for the random walk with a partially absorbing boundary at $x = \tilde{T}$. We are considering cases where $\tilde{\mathcal{P}}(y,0)$ is independent of $y$, so overall the purity picks up a multiplicative factor proportional to the total survival probability of the random walk. At long times, this scales as the return-probability of the random walk on an infinite chain~\cite{Redner2001}, resulting in $\tilde{\mathcal{P}}(x,t) \propto q^{-v_\text{E}(q)t} / \sqrt{t}$. 
We also provide independent confirmation of this result through large-scale simulations of random Clifford circuits (which exactly agree with the Haar-random circuit for the purity calculation), whose results are shown in Fig.~\ref{fig:Haar_num}.
Details on the method, as well as results on other interesting properties of this non-thermal volume-law entangled phase (from the point of view of dynamical purification~\cite{Gullans2020PRX}), are discussed in Appendix~\ref{app:purif}.

For the entropy calculation in the spacetime dual circuit, the relevant initial condition is one in which the domain wall sits right at the boundary, $x = \tilde{T}$, Fig.~\ref{fig:Haar}(c). 
The purity at time $t$ then becomes the purity of a subsystem in the long-time steady state with length $\tilde{\ell} = t$. Putting this all together, we find Eq.~\eqref{eq:HaarResult}. There are two noteworthy features of this result. First, the steady state exhibits volume-law entanglement, with a density $\lim_{\tilde{\ell}\to\infty} \left(\tilde{S}_{2,\infty}^{(a)} / \tilde{\ell}\right)$ equal to the speed of entanglement growth in the unitary dynamics, $v_\text{E}(q)$. 
Second, there is a subleading contribution, $(1/2)\log{\tilde{\ell}}$, that arises due to the boundary decoherence and is \emph{absent} in the original unitary circuit. In this sense, the long-time steady state is \emph{non-thermal}, despite its volume-law entanglement. 

As we noted, the $1/2$ coefficient of the logarithmic correction has its origins in the survival probability of the random walker decaying as $t^{-1/2}$. This is the appropriate quantity to calculate in the situation we have been considering so far, where the subsystem $A$ lives at the edge of the 1D chain (see Fig.~\ref{fig:boundary_dissip}(a)); this means that the domain wall has no preference on where to end up as long as it stays in the system. The calculation changes if we instead consider a subsystem in the bulk of an infinite chain. In that case, the boundary conditions in Fig.~\ref{fig:Haar}(c) need to be changed: we need to enforce the same domain wall configuration both at the initial and the final time on the partition function. This means that instead of the survival probability, we need to consider the \emph{return probability}\footnote{Since the two states, $\pm$ in the partition function are not orthogonal, the domain wall does not necessarily need to return exactly to its original location. Nevertheless, random walks ending up far away are exponentially suppressed, so the overall scaling is still that of the return probability.}. This instead decays as $t^{-3/2}$, giving rise to a correction $(3/2)\log{\tilde{\ell}}$ in the entropy.

We expect that the prefactors ($3/2$ in the bulk and $1/2$ at the edge) to be universal for spacetime-duals of chaotic unitary circuits. As can be seen from the above derivation, the origin of this prefactor lies in the diffusive dynamics of the purity domain wall. This description, in terms of a diffusing domain wall, is expected to apply for generic (space- and time translation invariant) chaotic unitary circuits in the long-time, large-distance limit (it is an emergent description on par with the usual diffusion of conserved quantities)~\cite{Jonay_2018,ZhouNahum_2020}. 
For this reason, our derivation above should also capture the universal features of spacetime duals of such chaotic circuits. The situation becomes more complicated for circuits that possess continuous symmetries. However, a similar description in terms of a diffusing domain wall exists even for these~\cite{Rakovszky_2019}, and therefore we conjecture that they would also exhibit the same $(3/2)\log{\tilde{\ell}}$ contribution. 
In fact, this term is exactly the same as the one conjectured to arise in hybrid (unitary-measurement) circuits~\cite{Li2019,Fan2020,LiFisher2020} in their volume-law phase; indeed, one could think of the domain wall picture we described as a microscopic realization of the `directed polymer in a random environment' (DPRE) effective description of Ref.~[\onlinecite{LiVijayFisher2021}]. This is further supported by a diagnostic discussed in Appendix~\ref{app:haar}, based on the scaling of mutual information between a single qubit and an extensive subsystem separated by a distance $x$. 
We find that this diagnostic scales as $\sim x^{-1.2}$ for duals of unitary circuits, in complete agreement with its behavior in generic hybrid circuits and the DPRE effective theory. 
These findings strongly suggest that duals of chaotic circuits realize the same non-thermal volume-law phase as generic monitored circuits. Given the superior analytical tractability of duals of unitary circuits, this analysis opens new routes for the study of this phase and its properties as an emergent error-correcting code~\cite{LiFisher2020}, which is an interesting direction for future research.

A notable exception to the universality of logarithmic corrections in duals of chaotic circuits is provided by dual-unitary circuits: in this case, the flipped circuit is itself unitary, and thus heats up to an infinite temperature state, rather than reaching non-thermal steady states of the kind realized by generic circuits; in particular, since the entropy is already maximal, there is no place for a logarithmic correction (negative log corrections are disallowed by subadditivity of the entropy). This can be understood from the operator spreading perspective discussed in Sec.~\ref{sec:boundary_decoh}. Dual-unitarity ensures~\cite{Prosen_entanglement,Prosen_correlations,Prosen_OpEnt} that operator strings always grow at the maximal possible speed. Consequently, there are no ``shrinking'' processes that could be killed by the boundary decoherence and the steady-state entanglement in the flipped direction is exactly equivalent to growth of entanglement in the unitary time direction, $\tilde{S}_\infty(\tilde{\ell}) = \tilde{\ell} = t$.

Finally, we need to comment on the issue of the order of averages in the Haar-random circuit. In the above, we focused on the annealed average, wherein we average the purity over circuit realizations before taking its logarithm. The quenched average, i.e. the average of $S_2$ itself, has a more complicated dynamics, described by the KPZ equation~\cite{Nahum2017,ZhouNahum_2019}. 
Among other things, this leads to a universal sub-leading contribution to the entropy growth in the unitary circuit of the form $a t^{1/3}$ (where the constant $a$ itself is non-universal). We expect that, upon flipping the circuit, this would lead to a contribution $a \tilde{\ell}^{1/3}$ in the quenched average of the entropy for the steady state, in agreement with recent results on random monitored Clifford circuits~\cite{LiVijayFisher2021}.
However, since the KPZ equation can be equivalently formulated in terms of multiple interacting random walkers, each of which would feel the same partially absorbing boundary, we conjecture that the overall result takes the form
\begin{equation*}
    \overline{\tilde{S}_2}(\tilde\ell) = v^{(q)}_E \tilde{\ell} + a \tilde{\ell}^{1/3} + \frac{1}{2} \log{\tilde{\ell}} + \text{const.} + \ldots
\end{equation*}
where we used $v^{(q)}_E$ to denote the speed associated to the quenched average of the entropy, which can be different from the one we calculated above for the annealed average.  Note that the $t^{1/3}$ contribution is a result of space-time randomness in the Haar circuit. As such, we expect the corresponding $\tilde{\ell}^{1/3}$ to be present in the spacetime duals of such random circuits, but to be absent for ones that are periodic in time. 


\section{Fractally entangled steady states\label{sec:fractal}}

\subsection{Griffiths circuit model}

Having examined the two extremes of dynamics in localized systems (Sec.~\ref{sec:anderson}) and in chaotic ones (Sec.~\ref{sec:Haar}), we now turn to slowly thermalizing dynamics in disordered systems, for instance near an MBL transition on the thermal side. 
There, entanglement growth is thought to be sub-ballistic, $S(t)\sim t^{\alpha}$, $0<\alpha<1$, due to the impact of Griffiths effects -- rare localized regions which serve as bottlenecks to the dynamics~\cite{Agarwal2015,BarLev2015,Vosk2015,Potter2015,Znidaric2016,Luitz_2016,LezamaSlow, NahumHuse_Griffith,Luitz_2019}. 
As both the localized and chaotic cases revealed a close relationship between $S(t)$ (entanglement growth in unitary circuits) on the one hand, and $\tilde{S}_\infty(\tilde\ell)$ (scaling in space of the late-time entanglement in the dual circuit) on the other, the sub-ballistic growth of $S(t)$ in these ``thermalizing Griffiths models'' suggests the intriguing possibility of producing late-time states with \emph{fractal} entanglement scaling, $\tilde{S}_\infty \sim \tilde{\ell}^\alpha$ with a tunable $\alpha \in (0,1)$.

\begin{figure}
    \centering
    \includegraphics[width=\columnwidth]{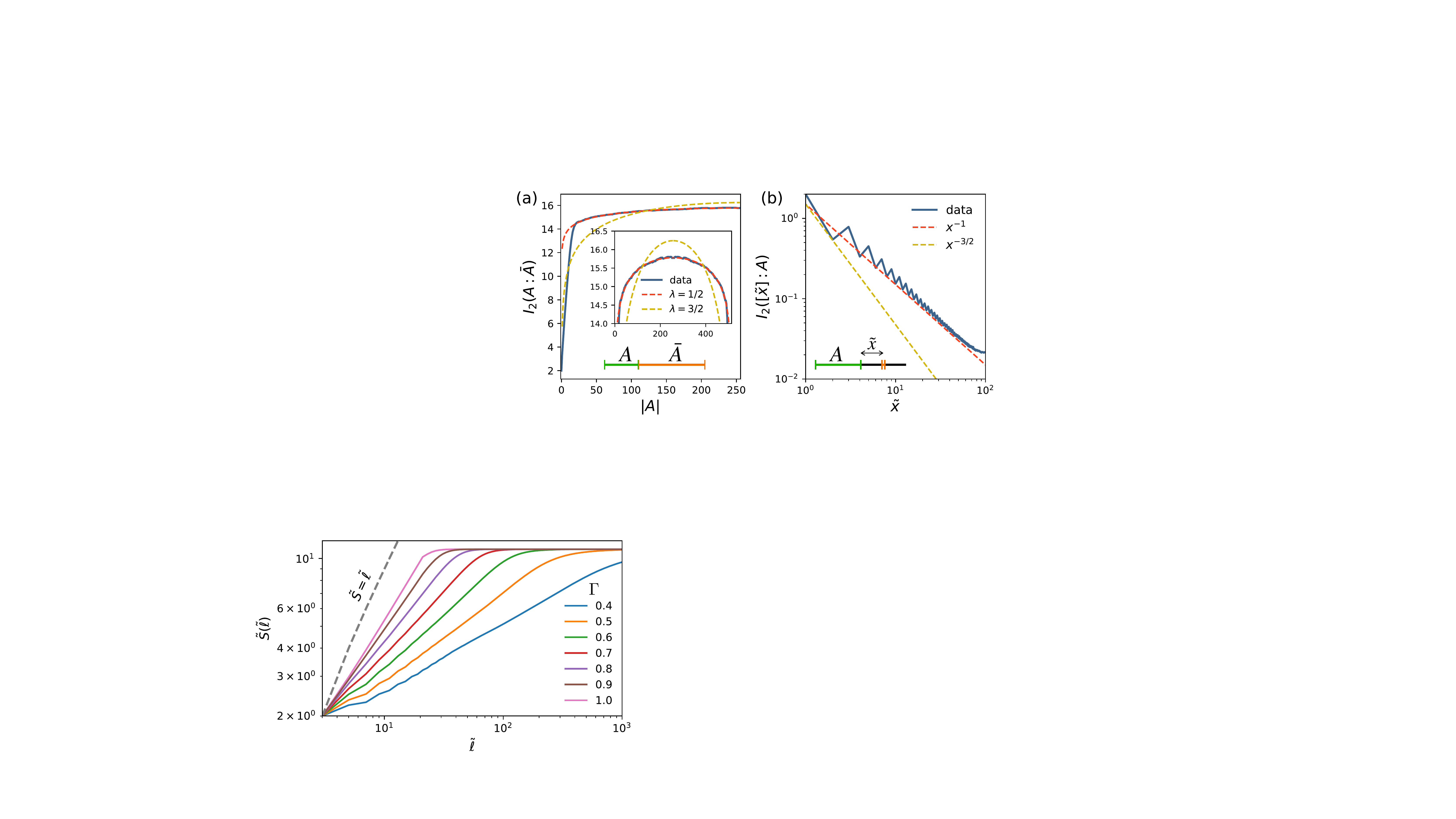}
    \caption{Entropy $\tilde{S}_\infty$ for the spacetime dual of the kicked Ising circuit Eq.~\eqref{eq:kicked_ising} in the ergodic phase, at depth $\tilde{T} = 11$ (numerical data from exact density matrix evolution for unitary evolution with edge decoherence). 
    The parameter $\Gamma$ controls the relative strength of disorder; the model has a thermal-to-MBL transition at $\Gamma_c\simeq 0.3$. 
    \label{fig:subballistic}}
\end{figure}

We see indications of this behavior in the kicked Ising model of Eq.~\eqref{eq:kicked_ising}, studied in the parameter regime of Ref.~[\onlinecite{Zhang2016}] (also summarized in Sec.~\ref{sec:models} and \ref{sec:anderson}) which features a transition from an MBL phase to a thermal one. 
As discussed in Sec.~\ref{sec:models}, this model has a parameter $\Gamma \in [0,1]$ which sets the disorder strength $W\propto \sqrt{1-\Gamma^2}$ and tunes across an MBL-to-thermal transition at $\Gamma_c\simeq 0.3$. Exact density matrix simulations for the unitary circuit with edge decoherence, whose results are shown in Fig.~\ref{fig:subballistic},  display a range of power law exponents that decrease closer to the MBL transition ($\Gamma_c\simeq 0.3$). However, the computational method limits us to modest depths $\tilde{T} \approx 10$, which in turn limits the available dynamic range and makes it difficult to characterize the fractal scaling. Likewise, the properties of the dual transition between logarithmic and fractal scaling of steady states are severely finite-size impacted. 

For this reason we turn to a more fruitful approach that relies on random circuits to capture coarse-grained features of the underlying subballistic dynamics. 
Here we use a family of models introduced in Ref.~[\onlinecite{NahumHuse_Griffith}], consisting of random circuits on one-dimensional spin chains where every bond $x$ is assigned a distinct rate $\gamma_x$, meant to capture the rate of entanglement propagation through a large disordered region in a coarse-grained way.
The rates are independently and identically sampled  from a distribution $P(\gamma) = (a+1)\gamma^a$, $\gamma\in[0,1]$, where the parameter $a\in (-1,+\infty)$ effectively controls the strength of disorder -- the smaller $a$, the more $P(\gamma)$ is concentrated near $\gamma=0$, the more ``weak links'' in the chain. 
Once rates $\{\gamma_x\}$ for all bonds are chosen, the system evolves via a spatiotemporally random circuit where each bond is acted upon by a gate every $\sim \gamma_x^{-1}$ time steps. Concretely, we use a brickwork circuit structure where each gate on bond $x$ is either $\eye$ (with probability $1-\gamma_x$) or a Haar-random gate (with probability $\gamma_x$), see Fig.~\ref{fig:fractal}(a).
In this model, entanglement grows \emph{sub-ballistically} as $S(t)\sim t^\alpha$, with $\alpha = (a+1)/(a+2) \in (0,1)$.~\cite{NahumHuse_Griffith}

Under spacetime duality, the disordered rates are not assigned to bonds in the chain, but rather to \emph{time steps}, thus creating ``temporal weak links'' in the evolution. 
At such a temporal weak link $\tilde{t}$, where $\gamma_{\tilde{t}}\ll 1$, the vast majority of bonds are acted upon by the spacetime dual of the identity, $\tilde{\eye} \propto \ket{B^+}\bra{B^+}$ -- i.e., they are projected onto Bell states. This results in a ``catastrophic'' event where most of the system becomes disentangled, leaving behind a sparse network of un-measured entangled sites.
The repetition of these events over all scales (in time $\tilde t$ and rate $\gamma$) prevents entanglement from ever saturating to a volume-law, and instead gives rise to a \emph{fractal} pattern of entanglement in late-time states.

\subsection{Fractal entanglement}

This result is straightforwardly proved in the edge decoherence picture by adapting the discussion in Ref.~[\onlinecite{NahumHuse_Griffith}], with minor changes from the closed-system case to the edge-decoherence one. 
First, imagine a situation where the rates are uniform, $\gamma_x\equiv \gamma^{(0)}$, except for a single ``weak link'' at a distance $x_1$ from the decohering edge, where the rate is $\gamma_{x_1}\equiv \gamma^{(1)} \ll \gamma^{(0)}$. 
At early times, the depolarizing noise quickly decoheres the region between the weak link and the boundary, giving a ballistic growth $S(t)\sim \gamma t$ similar to the Haar random case discussed in Sec.~\ref{sec:Haar}. At later times, the bottleneck at $x_1$ sets the rate of entropy growth\footnote{Here, we rely on the subadditivity of the von Neumann entropy: the entropy of $\rho_B$ is upper bounded by the sum of the entropy of the cut at $x_1$ and the entropy of the region between $x_1$ and the boundary.},
while the region between the weak link and the boundary essentially saturates to a fully mixed state, resulting in $S(t) = \min\{\gamma^{(0)} t, x_1 + \gamma^{(1)} t \}$. 
Generalizing this argument to the case with a hierarchy of (increasingly far, increasingly weak) links at distances $x_n$ (with $x_0=0$) and rates $\gamma^{(n)}$ gives 
\begin{equation*}
S(t) = \min_n \{x_n + \gamma^{(n)} t\} \;.    
\end{equation*} 
In other words, at any given time scale $t$, there is one ``dominant'' bottleneck (the value of $n$ minimizing the argument) that is mainly responsible for slowing down the spreading of decoherence through the system. 
Now, given a length scale $\zeta$, the typical value of the slowest rate encountered within that lengthscale is~\cite{NahumHuse_Griffith}
$\gamma_\text{min}^\text{(typ)}(\zeta) \sim \zeta^{-1/(a+1)}$. 
Therefore we conclude that, typically, 
\begin{align}\label{eq:fractal_exp}
S(t)\sim \min_\zeta \{ \zeta + \zeta^{-1/(a+1)} t\} \sim t^\alpha,
& &
\alpha = \frac{a+1}{a+2}\;,
\end{align}
i.e. the same sub-ballistic scaling as in Ref.~[\onlinecite{NahumHuse_Griffith}].
We emphasize that this is the leading-order scaling, and we will not focus on subleading corrections at this level.

\begin{figure}
    \centering
    \includegraphics[width=\columnwidth]{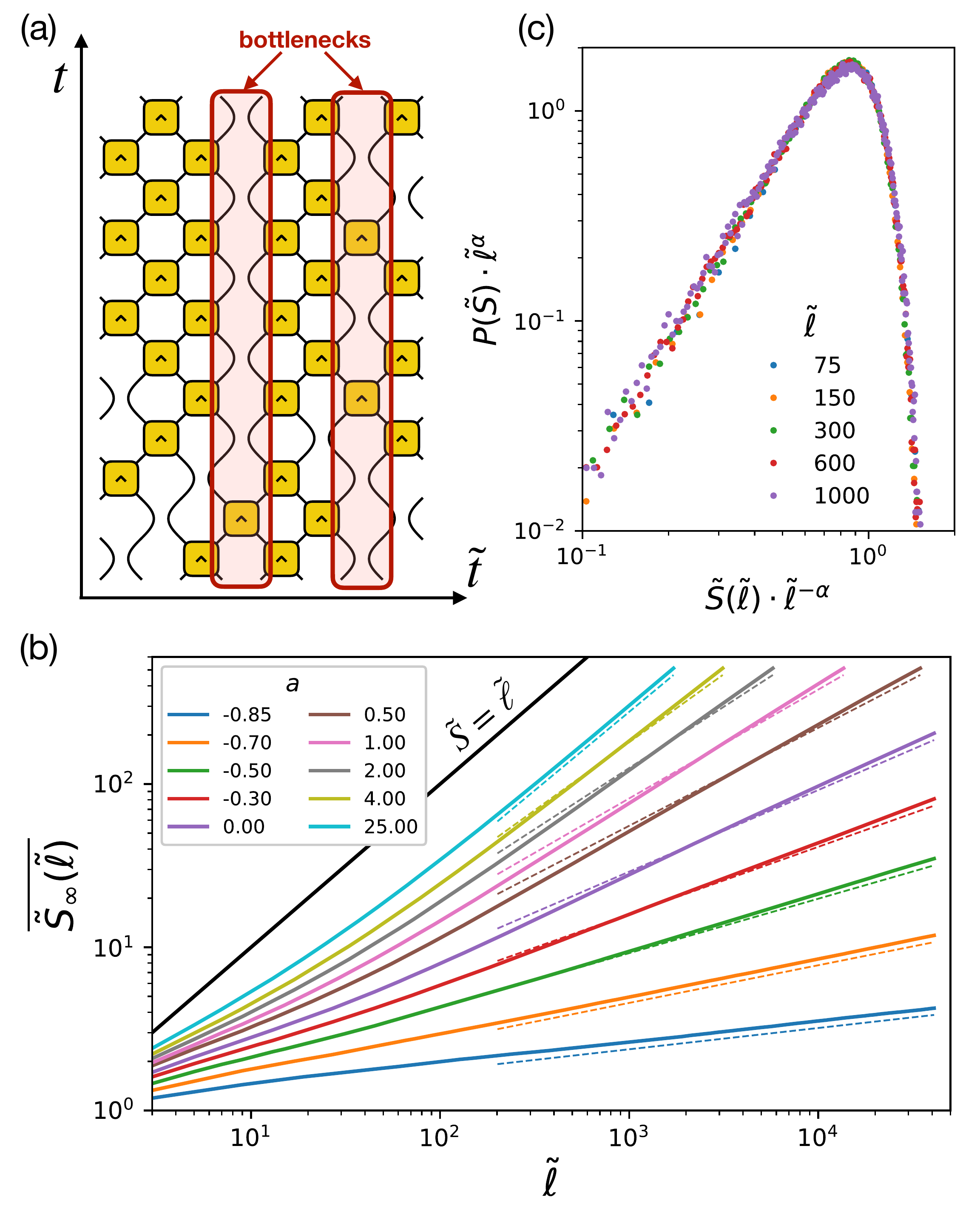}
    \caption{Fractal entanglement from Griffiths circuit model. 
    (a) Schematic of thecircuit: each bond $x$ is assigned a rate $\gamma_x$ that sets how often unitary gates act there; bonds where $\gamma_x\ll 1$ act as bottlenecks for the spread of entanglement (highlighted). Under spacetime duality, bottlenecks correspond to ``catastrophic events'', where simultaneous Bell measurements are performed on most qudit pairs. (b) Average entanglement entropy $\overline{\tilde{S}(\tilde\ell)}$ in late-time output states of a version of this circuit where non-identity gates are drawn from the 2-site Clifford group (using $\tilde{T} = 1024$ qubits and averaging over $10^5$ disorder realizations). By variying the effective disorder strength $a$, $\overline{\tilde{S}}$ can be made to scale as ${\tilde\ell}^\alpha$ with any power-law exponent $0<\alpha<1$. The black line represents the maximal entropy $\tilde{S}=\tilde{\ell}$; dashed lines represent the predicted power-law scaling $\alpha = (a+1)/(a+2)$. The disagreement for intermediate $a$ is due to subleading corrections (see Appendix~\ref{app:griffiths_rw}).
    (c) The entire distribution of $\tilde{S}$ (not just its mean) shows a scaling collapse, $P(\tilde{S}) \sim \tilde{\ell}^{-\alpha} P(\tilde{S}/\tilde{\ell}^{\alpha})$ (data for $a=1$, fit exponent $\alpha=0.73$ vs predicted $2/3$).
    \label{fig:fractal}}
\end{figure}

We confirm this result with numerical simulations, using two methods: (i) stabilizer numerical simulations, where the non-identity gates are sampled uniformly from the Clifford group on qubits ($q=2$); and (ii) a recursive (random walk) construction like the one used in Sec.~\ref{sec:Haar}, but adapted to the presence of random rates as detailed in Appendix~\ref{app:griffiths_rw}. 
In both cases, we simulate unitary dynamics with edge decoherence. 
The results of both methods confirm the unconventional scaling of the averaged entanglement entropy with subsystem size, $\tilde{S}_\infty \sim {\tilde\ell}^\alpha$ with fractional exponents $0<\alpha<1$; we show this in Fig.~\ref{fig:fractal}(b) for the Clifford simulations and in App~\ref{app:griffiths_rw} for the random walk method. Small numerical discrepancies between the exponent predicted by Eq.~\eqref{eq:fractal_exp} and the data (most evident at intermediate $a$) arise from large subleading corrections to scaling, which are expected to be present already in the unitary case, as we discuss in Appendix~\ref{app:griffiths_rw}.
Beyond the fractional power-law scaling of the mean entropy, we find that all moments of the entropy distribution scale in the same way, as manifested by the collapse of the probability distribution $P(\tilde{S}_\infty)$ onto a scaling ansatz $P(\tilde{S}_\infty) \propto {\tilde\ell}^{-\alpha} f(\tilde{S}_\infty / {\tilde\ell}^{\alpha})$, shown in Fig.~\ref{fig:fractal}(c).
This means that the spatial entanglement profile of a typical state drawn from this ensemble appears \emph{statistically self-similar} over all length scales, justifying the \emph{``fractal''} label. This is in contrast with, e.g., the volume-law phase of Sec.~\ref{sec:Haar} where one expects (under quenched averaging) $\overline{S} \sim \ell$ and $\delta S \sim \ell^{1/3}$ (owing to the conjectured KPZ scaling): the relative size of fluctuations in this case is scale-dependent and not self-similar.

Another intriguing aspect of these models is studied in Appendix~\ref{app:purif}, where we show data on their purification dynamics~\cite{Gullans2020PRX}. We find that they realize a family of novel \emph{critical purification phases} with continuously variable dynamical exponent.
In particular, the entropy of a maximally mixed initial state decays as $\tilde{S}\sim \tilde{L} / \tilde{t}^{1/z_p}$, with $z_p = a+1$, and at late time crosses over to an exponential decay (also as a function of the ratio $\tilde{t} / \tilde{L}^{z_p}$), suggestive of CFT behavior~\cite{Li2020}.

The $\tilde{\ell}^\alpha$ scaling derived above captures the leading order behavior in this model. Even in the unitary case, there are subleading corrections, as we noted above. In the spacetime-dual circuit we expect that further logarithmic corrections should appear due to the edge decoherence, just as they did in both the localized and chaotic models studied in Sections~\ref{sec:anderson} and~\ref{sec:Haar}.
The most straightforward argument for these comes from generalizing the random walk calculation of Sec.~\ref{sec:Haar} to the present model. In this modified random walk, we now have locations in the circuit where the Haar-random gate is replaced by the identity: these have no effect on the ``purity domain wall'', while the partially absorbing boundary condition is unchanged. Therefore, the purity should again pick up a power-law decaying contribution from the survival/return probability of the random walker (depending on whether we are considering a subsystem near the edge or in the bulk). 
However, the power of the decay might be different from the ones derived in Sec.~\ref{sec:Haar}. In that case, the endpoint of the domain wall had diffusive dynamics, a fact that is closely related to the diffusive broadening of operator wavefronts~\cite{Nahum2018}. This latter feature is also expected to change in the Griffiths phase, with a separate, continuously changing ``front broadening exponent'' at strong enough disorder~\cite{NahumHuse_Griffith}. Consequently, we expect that the prefactor of the logarithmic correction to the entropy also becomes a continuously varying parameter deep in the Griffiths regime.

\subsection{Robustness to breaking of unitarity}

We derived these fractally-entangled phases by dualizing unitary circuits $U$ with subballistic entanglement growth. 
It is natural to ask whether or not this duality is a necessary condition, i.e. how robust these phases are to perturbing the ``original'' circuit $U$ away from unitarity.
As we have shown, area-law phases are generically ruled out in non-unitary circuits whose dual is unitary; but upon lifting this constraint, area-law phases are expected to appear again when a (suitably defined) ``measurement rate'' is high enough. 
Do fractally entangled phases survive then, at least in parts of parameter space for sufficiently weak perturbations? Or are they immediately wiped out when unitarity of $U$ is broken?

To address this question, we perturb the unitary ``Griffiths circuit'' of Fig.~\ref{fig:fractal}(a) away from unitarity, and study the effect of this perturbation on entanglement in the dual circuit.
Concretely, we consider a unitary-measurement circuit with random unitary gates and two-qubit Bell measurements, where the measurement rate is time-dependent, $p(t) \equiv 1-\gamma_t$, with $\gamma$ sampled from $P(\gamma) = (a+1)\gamma^a$ independently at each time step.
This circuit is generally non-unitary in both time directions.
However, when the gates are restricted to be dual-unitary, this circuit becomes unitary in the space direction: namely it is dual to a unitary ``Griffiths circuit'' as in Fig.~\ref{fig:fractal}, where all the non-identity gates are dual-unitary. 
It is possible to interpolate smoothly between the two cases (fully generic vs dual-unitary gates) by tuning the distance of the gate set from dual-unitarity.
This is particularly straightforward for the two-qubit Clifford group, which is discrete and such that exactly half of the elements are dual-unitary.
Sampling preferentially from the dual-unitary half of the group, say with probability $1-\delta$, gives a parameter $\delta$ that quantifies the breaking of unitarity in the space direction.
Do the fractal phases we found at $\delta = 0$ survive to finite $\delta>0$?

\begin{figure}
    \centering
    \includegraphics[width=\columnwidth]{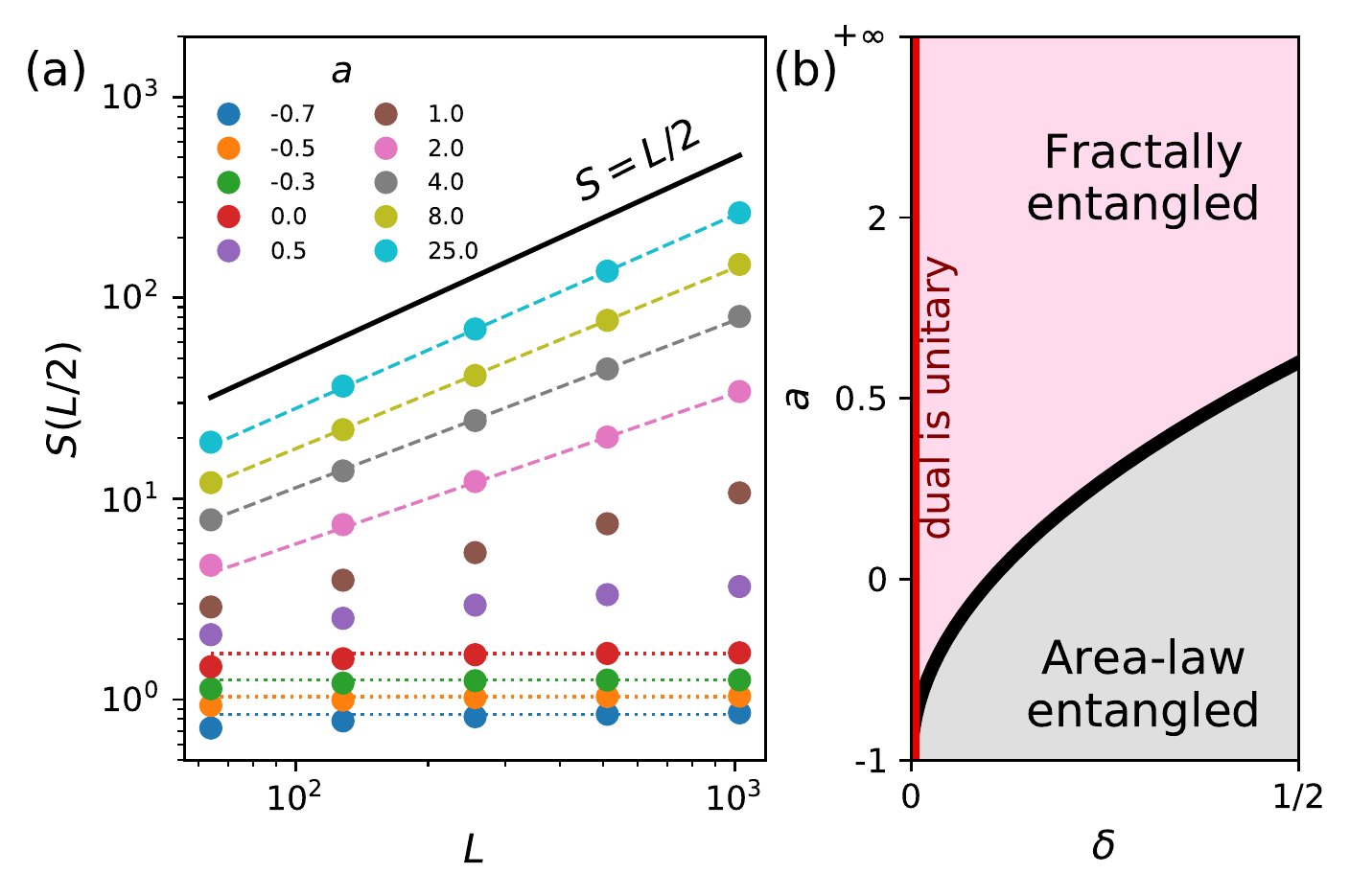}
    \caption{Robustness of fractal entanglement in $\tilde{U}$ to the breaking of unitarity in $U$. 
    (a) Half-cut entanglement entropy $S(L/2)$ for a pure state evolving under random Clifford gates and Bell measurements; the measurement probability $1-\gamma_t$ depends randomly on time, with $\gamma$ drawn from $P(\gamma) \propto \gamma^a$ independently at each time step. 
    Data from stabilizer simulations, with $L<T<2L$, averaged over between $10^4$ and $5\times 10^5$ realizations depending on size. The system enters an area-law phase for $a\lesssim 0$ (dotted lines are constants), while fractal entanglement persists at $a\gtrsim 1$ (dashed lines have slope $\alpha \equiv (a+1)/(a+2)$). 
    (b) Sketch of conjectured phase diagram as a function of the bottleneck parameter $a$ and the deviation from unitarity $\delta$ (see text; panel (a) corresponds to $\delta=1/2$). The $\delta=0$ line corresponds to the circuit $U$ being unitary, which forbids the area-law phase in $\tilde{U}$.}
    \label{fig:robustness}
\end{figure}

The steady-state entanglement in this family of unitary-measurement circuits, shown in Fig.~\ref{fig:robustness}(a) for $\delta = 1/2$ (i.e. uniform sampling of Clifford gates), reveals an interesting picture.
As anticipated, an area-law phase appears:
the average rate of measurements in this circuit is $\overline{p} = 1 - \overline{\gamma} = \frac{1}{a+2}$, which becomes large at small $a$ (going to 1 as $a \to -1$); absent obstructions, this is expected to lead to an area-law phase. 
Indeed, that is what we see for $a\lesssim 0$.
However, upon increasing $a$, the system transitions out of the area-law phase and into a family of fractally entangled phases.
We verify that for $a\geq 2$ the scaling is consistent with the expected one, $S(\ell) \sim \ell^\alpha$, $\alpha = \frac{a+1}{a+2}$ (dashed lines in Fig.~\ref{fig:robustness}(a)).
At intermediate $a$, the data is suggestive of a logarithmically-entangled critical point (at $a_c\approx 0.5$) that gives way to fractal entanglement with exponent $\alpha(a)$ continuously varying from $\alpha(a_c) = 0$ to $\alpha(a)\simeq \frac{a+1}{a+2}$ at large $a$. While a detailed study of this transition is left for future research, it is clear that fractal entanglement does indeed survive in a large part of parameter space, even when the ``original'' circuit $U$ is strongly perturbed away from unitarity.

A conjectured sketch of the phase diagram as a function of the ``bottleneck parameter'' $a$ and the unitarity-breaking perturbation strength $\delta$ is shown in Fig.~\ref{fig:robustness}(b). 
The area-law phase in $\tilde{U}$ appears continuously as the circuit $U$ is detuned from unitarity, while the phases identified at $\delta = 0$ survive in part of parameter space.
We conjecture that this behavior is generic -- i.e. that all the entanglement phases identified in this work should be robust to weak breaking of unitarity in the transverse direction.

\section{Summary and Outlook \label{sec:conclusion}}

In this work, we have explored entanglement in non-unitary circuits that are spacetime duals of local unitary circuits~\cite{Ippoliti2021PRL}.
Focusing on this class of non-unitary circuits has allowed us to translate the rich variety in entanglement growth displayed by unitary circuits into an equally rich variety of steady-state phases produced by non-unitary dynamics. 
Most notably, these include a large and robust family of \emph{fractally entangled} late-time states, whose entanglement evades the usual categories of volume-, area-, or log-law that cover nearly the entirety of eigenstates or steady-states encountered in unitary dynamics.

Another important advantage of these circuits lies in their analytical tractability: ideas and methods developed for unitary dynamics can be borrowed and used to great effect, providing an analytical handle on non-unitary dynamics away from well-understood limits (e.g. large local Hilbert space, etc.)
In particular, we have derived a general identity relating steady-state entanglement in these non-unitary circuits to the entropy of mixed states evolving through a combination of unitary dynamics and decoherence.
This powerful mapping has allowed us to identify and characterized logarithmic corrections to entanglement in various regimes:
in the duals of chaotic circuits, these corrections sharply distinguish the volume-law entangled phase from a thermal one (while possibly placing it in the same universality as the non-thermal volume-law phase in weakly monitored circuits);
in the duals of localized circuits, these corrections can even become the \emph{leading} contributions, essentially ruling out area-law entangled phases.

We also detailed how these novel steady states can be prepared experimentally on near-term quantum simulators, with an experimental protocol based on quantum teleportation that relies on ordinary unitary evolution and a \emph{vanishing density} of postselected measurements in spacetime (as opposed to the finite density that characterizes generic unitary-measurement circuits). 
Lowering the required number of measurements has important implications for the practicality of reliably preparing these states without the use of auxiliary classical simulation.

Our work raises several intriguing questions for future research.
Understanding more about the nature of the fractally-entangled steady states we found in the duals of ``thermalizing Griffiths'' phases, and what other contexts they might occur in, is an exciting goal for future research.
It is particularly intriguing to speculate about the possibility of realizing these unconventional states (which are natively the output of $1+1$-dimensional non-unitary circuits) as ground states of generalized (possibly non-Hermitian) Hamiltonians, and study the role of symmetry therein~\cite{Bao2021}.

Another interesting open question is the precise nature of the volume-law phase we identified in the spacetime duals of generic clean (or weakly disordered) unitary circuits. As we noted, the universal prefactor of the logarithmic correction agrees with what is expected in weakly-monitored unitary circuits. Another diagnostic based on mutual information (analyzed in Appendix~\ref{app:haar}) does not conclusively discriminate the two phases, either. This leaves two interesting possibilities: (i) that the duals of chaotic unitary circuits realize a novel non-thermal volume-law entangled phase; (ii) that they provide an exact realization of the same phase in a way that is substantially more tractable in theory and more accessible in experiment. 
Determining which of these two scenarios is realized is an exciting question for future research.

More generally, it would be interesting to build a more complete understanding of these non-thermal states and the place they occupy in the broader landscape of dynamical many-body states, e.g. through the lens of chaos, complexity and thermalization. (``Non-thermal'' is, after all, a rather vague label). It would also be interesting to study the properties of non-unitary phase transitions dual to unitary transitions; the Floquet Ising model studied here does have a (unitary) transition between MBL and thermalizing phases, but the properties of its dual non-unitary counterpart were challenging to study due to the limitations of finite system size numerics. 

In this work we have mainly focused on quantum entanglement; however, characterizing these novel steady states from the standpoint of quantum order is another natural direction for future research. Can these states host nonequilibrium ordered phases? In most known cases in the unitary domain, out-of-equilibrium phases rely on MBL and \emph{eigenstate order}~\cite{lpqo, Khemani2016}; how does the impossibility of area-law states in these spacetime-dual circuits affect the definition of phases? And how are the (leading or sub-leading) logarithmic corrections manifested in correlation functions? Extending these ideas to higher dimension may also open up exciting directions in relation to topology, e.g. by considering the spacetime duals of $2+1$-dimensional Floquet topological phases~\cite{Po2016}. 

Looking ahead, perhaps the most exciting challenge is to push the ``non-unitary frontier'' by identifying other interesting (and experimentally relevant) corners of non-unitary evolutions that may further reveal new phenomena and broaden our understanding of quantum dynamics.

\acknowledgements We thank David Huse, Sarang Gopalakrishnan and Michael Gullans for many insightful discussions and collaboration on related topics. We also acknowledge helpful discussions with Ehud Altman, Yimu Bao, Soonwon Choi, Yaodong Li, Adam Nahum and Chaitanya Murthy. This work was supported with funding from the Defense Advanced Research Projects Agency (DARPA) via the DRINQS program (M.I.), the Sloan Foundation through a Sloan Research Fellowship (V.K.) and by the US Department of Energy, Office of Science, Basic Energy Sciences, under Early Career Award No. DE-SC0021111 (V.K.). The views, opinions and/or findings expressed are those of the authors and should not be interpreted as representing the official views or policies of the Department of Defense or the U.S. Government. M.I. was funded in part by the Gordon and Betty Moore Foundation's EPiQS Initiative
through Grant GBMF4302 and GBMF8686. T.R. is supported by the Stanford Q-Farm Bloch Postdoctoral Fellowship in Quantum Science and Engineering. 
Numerical simulations were performed on Stanford Research Computing Center's Sherlock cluster.

\emph{Note Added:} As we were finishing this manuscript, we became aware of related work ~\cite{TarunFlipped} (appearing in the same arXiv posting) which also considers entanglement dynamics in spacetime duals of unitary circuits. Our results agree where they overlap, but our works are largely complementary. 

\bibliographystyle{unsrt}
\bibliography{flip.bib}


\appendix


\section{Details on edge decoherence \label{app:edge_d}}

Here we clarify some technical details on the mapping to edge decoherence introduced in Sec.~\ref{sec:setup}, and some results that follow from it. 
We have argued that, taking the limit $\tilde{L}\to\infty$ with $\tilde{\ell}$ finite (i.e. when $A$ is a finite subsystem of a semi-infinite chain), the entire $\bar{A}$ section of the circuit in Fig.~\ref{fig:boundary_dissip}(a) can be elided (for the purpose of computing the entropy of $A$), and it suffices to consider the mixed state $\rho_B$ produced at the spacelike cut $B$ between the timelike surfaces $A$ and $\bar{A}$. 
Here we unravel the argument in more detail, and in the process clarify how big $\tilde{L}$ needs to be in various models and what the fine-tuned exceptions to this result look like.
Finally, as a corollary, we show that area-law entangled steady states are impossible (again up to trivial, fine-tuned exceptions).

\subsection{Uniqueness of the steady state under edge decoherence}

To begin, we consider the layout of Fig.~\ref{fig:boundary_dissip}(c) -- unitary evolution with edge decoherence, for a variable time $t$ (also identified with subsystem size $\tilde\ell$ in the ``flipped'' circuit).
Let $\rho_B(t)$ be the mixed state produced by $t$ layers of unitary gates accompanied by an erasure channel on the last qubit, 
\begin{equation}
    \rho_B(t)\equiv \Phi_\text{edge} \circ \mathcal{U}_t \circ \Phi_\text{edge} \circ \cdots \mathcal{U}_1 [\rho_B(0)] \;,
    \label{eq:edge_depo_channel}
\end{equation}
where $\mathcal{U}_\tau[\rho] \equiv U_\tau \rho U_\tau^\dagger$ describes the application of one layer of unitary gates at time $\tau$ (we do not assume a Floquet circuit) and $\Phi_\text{edge}[\rho] \equiv \Tr_\text{edge}(\rho)\otimes \eye_\text{edge}/q$ is the erasure (or fully depolarizing) channel acting on the edge qubit. 
We will prove the following statement:
{\it up to fine-tuned exceptions in the circuit choice $\mathcal{U}_\tau$, the quantum channel Eq.~\eqref{eq:edge_depo_channel} maps all input states to the maximally mixed state $\rho_B\equiv \eye^{\otimes L}/q^L$ as $t\to\infty$.}

The proof can be succinctly stated in the operator spreading language. First, let us consider time-periodic dynamics, $\mathcal{U}_\tau = \mathcal{U}$. We imagine expanding our prospective steady state, $\rho_\infty$ in a basis of Pauli strings. $\Phi_\text{edge}$ is a projector onto the subspace spanned by strings that act as the identity on the edge site. Clearly, $\rho_\infty$ must belong to this subspace: $\Phi_\text{edge}[\rho_\infty] = \rho_\infty$. Moreover, it also cannot develop any component on the orthogonal subspace (i.e., the space of operators that do \emph{not} act as the identity on the edge), since such a component would be killed by the projection, decreasing the norm of $\rho_\infty$; therefore $(\Phi_\text{edge}\circ \mathcal{U})[\rho_\infty] = \mathcal{U}[\rho_\infty]$. In other words, we are looking for \emph{exactly} conserved operators of the Floquet unitary, $\mathcal{U}[\rho_\infty] = \rho_\infty$ that have \emph{no} support on the edge; obviously this should only occur in highly contrived cases. While this argument works only for Floquet systems, where steady states can exist in a strict sense, it easily generalizes to time-dependent circuits as well: the only way for an initial state to maintain its Frobenius norm (i.e., its total purity) forever is if it \emph{never} develops a component acting non-trivially on the edge.

Exceptions where the above argument fails are circuits with exact ``blockades'':
unitaries where some operators are strictly confined to a segment of the circuit not containing the edge site.
The simplest example is a system that is split into decoupled subsystems at a missing bond, where $U = \eye^{\otimes 2}$ at all times. A somewhat less trivial example is given by the Clifford toy model of MBL discussed in Ref.~[\onlinecite{ChandranLaumann_2015}] which exhibits l-bits \emph{exactly} localized on certain sites.
Intuitively, one can understand this in the language of operator hydrodynamics~\cite{Keyserlingk2018,Nahum2018,Khemani_2018,Rakovszky_2018}: the edge site functions as a ``sink'' for the operator weight which is otherwise conserved by the unitary dynamics; the only way to avoid all the weight leaving at the sink is through a strict blockade that prevents some operators from reaching the sink.
Barring these fine-tuned ``blockaded'' circuits, the dynamics with edge decoherence Eq.~\eqref{eq:edge_depo_channel} always converges to the maximally mixed state. The time scale for this process, that we will dub the ``decoherence time'' $t_d$, is closely related to entanglement dynamics in the unitary circuit. In localized models, as we show in Sec.~\ref{sec:anderson}, the system takes an exponentially long time $t_d \sim \exp(L)$ to fully decohere. On the opposite end, chaotic systems will take time $t_d \sim L$.


\subsection{Normalization of the wavefunction}

With this result in hand, we can show that the norm of a wavefunction $\ket{\psi(\tilde{t})}$ evolving under the non-unitary spacetime-dual circuit is statistically conserved (and exactly conserved in the thermodynamic limit $\tilde{L}\to\infty$), even though the individual gates $\tilde{U}$ are not norm-preserving. 

We assume the initial state $\ket{\psi(0)}$ is a product state of Bell pairs, $\ket{B^+}^{\otimes \tilde{L}/2}$, as in Fig.~\ref{fig:boundary_dissip}(a).
Then, we aim to calculate the squared norm $\braket{\psi(\tilde t)|\psi(\tilde{t})}$. 
The tensor network expressing this quantity is exactly the same as in Fig.~\ref{fig:boundary_dissip}(a), with the traced-out subsystem $A$ encompassing the whole system. 
This can be interpreted from bottom to top as (i) an initial state $\ket{B^+}^{\otimes \tilde{t}/2}$ (ii) evolved with unitary gates and edge decoherence for $\tilde{\ell}$ time steps until (iii) all qubits are projected onto $\bra{B^+}^{\otimes \tilde{t}/2}$. 
However, some care must be taken in the bookkeeping of normalizations when converting between the unitary and non-unitary ``arrows of time''.
In the non-unitary direction the tensor network comes with a prefactor of $q^{-\tilde{\ell}/2}$: $1/q$ for every Bell-pair initial state $\ket{B^+}\bra{B^+}$ (left edge).
In the unitary direction, it comes with a prefactor of $q^{-\tilde{\ell}/2 - \tilde{t}}$: $1/q$ for every action of the erasure channel at the right edge, and $1/q$ for every Bell-pair in either the initial (bottom) or final (top) Bell states. 
The net result is a mismatch by $q^{-\tilde{t}}$, which must be taken into account when ``flipping'' the circuit.

For large systems (with $\tilde{\ell}\gg t_d$), the associated unitary-decoherence dynamics reaches the fully-mixed steady state, by the very definition of the decoherence time $t_d$; 
therefore, projecting onto the final Bell-pair product state (or any other state!) yields a fixed amplitude of 
\begin{align*}
    \Tr\left[\left(\ket{B^+}\bra{B^+}\right)^{\otimes \tilde t/2} \rho_\text{out}\right] & = \bra{B^+}^{\otimes \tilde{t}/2} \left( \frac{\eye}{q}\right)^{\otimes \tilde{t}} \ket{B^+}^{\otimes \tilde{t}/2} \\
    & = q^{-\tilde{t}} \;.
\end{align*}
This factor gets exactly cancelled when converting back to the non-unitary arrow of time, so that $\braket{\psi(\tilde{t})|\psi(\tilde{t})} = 1$. 

\subsection{Entanglement calculation}

We now consider the setup of Fig.~\ref{fig:boundary_dissip}(a) and show that, for the purpose of computing the entropy, it reduces to that of Fig.~\ref{fig:boundary_dissip}(b) in the limit of $\tilde{L}\to\infty$. 
We do so by splitting the tensor network that expresses the output state $\ket{\psi}$ into two pieces along the entanglement cut $B$:
\begin{equation}
    \psi_{a\bar{a}} \equiv \sum_{b} \phi_{ab} \chi_{b\bar{a}} \;,
    \label{eq:app_psidecomposition}
\end{equation}
where $a$ and $\bar{a}$ index basis states for timelike subsystems $A$ and $\bar{A}$ while $b$ indexes basis states for the spacelike subsystem $B$ at the entanglement cut (consisting of $\tilde{t}$ qubits). 
$\phi$ and $\chi$ thus represent the two parts of the tensor network that, when glued together at $B$ as in Eq.~\eqref{eq:app_psidecomposition}, yield the output wavefunction on $A\cup\bar{A}$. 

Now, we aim to compute the entropy of the reduced density matrix on $A$:
\begin{align}
    (\rho_A)_{aa'}
    \equiv \sum_{\bar{a}} \psi_{a\bar{a}} \psi^\ast_{a'\bar{a}}
    = \sum_{b,b'} \phi_{ab} \phi^\ast_{a'b'} 
    \left( \sum_{\bar{a}} \chi_{b\bar{a}} \chi^\ast_{b'\bar{a}} \right). \nonumber
\end{align}
The sum in parentheses represents the reduced density matrix on $B$ produced from running unitary dynamics with edge decoherence for a time $\tilde{L}-\tilde{\ell}$. 
But if $\tilde L$ is taken to be much bigger than the decoherence time $t_d$, then this density matrix is generically the identity: $\sum_{\bar{a}} \chi_{b\bar{a}} \chi^\ast_{b'\bar{a}} = \delta_{bb'}$ (normalizations are dealt with as above).
Therefore we have
\begin{equation}
    \rho_A = \sum_{a,a',b} \phi_{ab} \phi^\ast_{a'b} \ket{a}_A \bra{a'}_A \;.
\end{equation}
Finally, we can purify $\rho_A$ to a wavefunction on $A\cup B$, $\ket{\phi}_{AB} \equiv \phi_{ab} \ket{a}_A\otimes \ket{b}_B$ (this is the tensor network in Fig.~\ref{fig:boundary_dissip}(b)), 
and recover the same entropy by tracing out $A$ instead:
\begin{equation}
    \rho_B = \sum_{a,b,b'} \phi_{ab} \phi^\ast_{ab'} \ket{b}_B \bra{b'}_B \;.
    \label{eq:app_edgedepo_result}
\end{equation}
The density matrix $\rho_B$ is precisely the output of $\tilde{\ell}$ layers of unitary evolution and edge decoherence on $\tilde{T}$ qubits, as in Fig.~\ref{fig:boundary_dissip}(c) in the main text.

\subsection{Impossibility of area-law steady states}

Equipped with the results above, we can now rule out the existence of area-law entangled late-time states in all but a highly fine-tuned set of circuits. 
Having $\ket{\psi(\tilde t)}$ be area-law entangled would mean that the entropy of subsystem $A$ (of size $\tilde{\ell}$) should \emph{not} grow with $\tilde{\ell}$ past a finite amount (independent of $\tilde{T}$ after an initial transient). 
But in the edge-decoherence picture of Eq.~\eqref{eq:app_edgedepo_result} and Fig.~\ref{fig:boundary_dissip}(c), this means that the mixed state $\rho_B(\tilde{\ell})$ should reach a \emph{finite} entropy at arbitrarily large $\tilde{\ell}$. 
This would rule out the maximally mixed steady state $\rho_\infty = (\eye/q)^{\otimes \tilde{t}}$ (whose entropy is $\tilde{t}$). 
As we saw above, this is impossible in all but a set of fine-tuned ``blockaded'' circuits. This explains the absence of a ``pure phase'' in the model studied in Ref.~[\onlinecite{Ippoliti2021PRL}], except for the trivial circuit consisting of $\eye$ only (which indeed is trivially ``blockaded'' in the above sense).


\section{Dynamical purification in spacetime-dual circuits \label{app:purif}}

\begin{figure}
    \centering
    \includegraphics[width=\columnwidth]{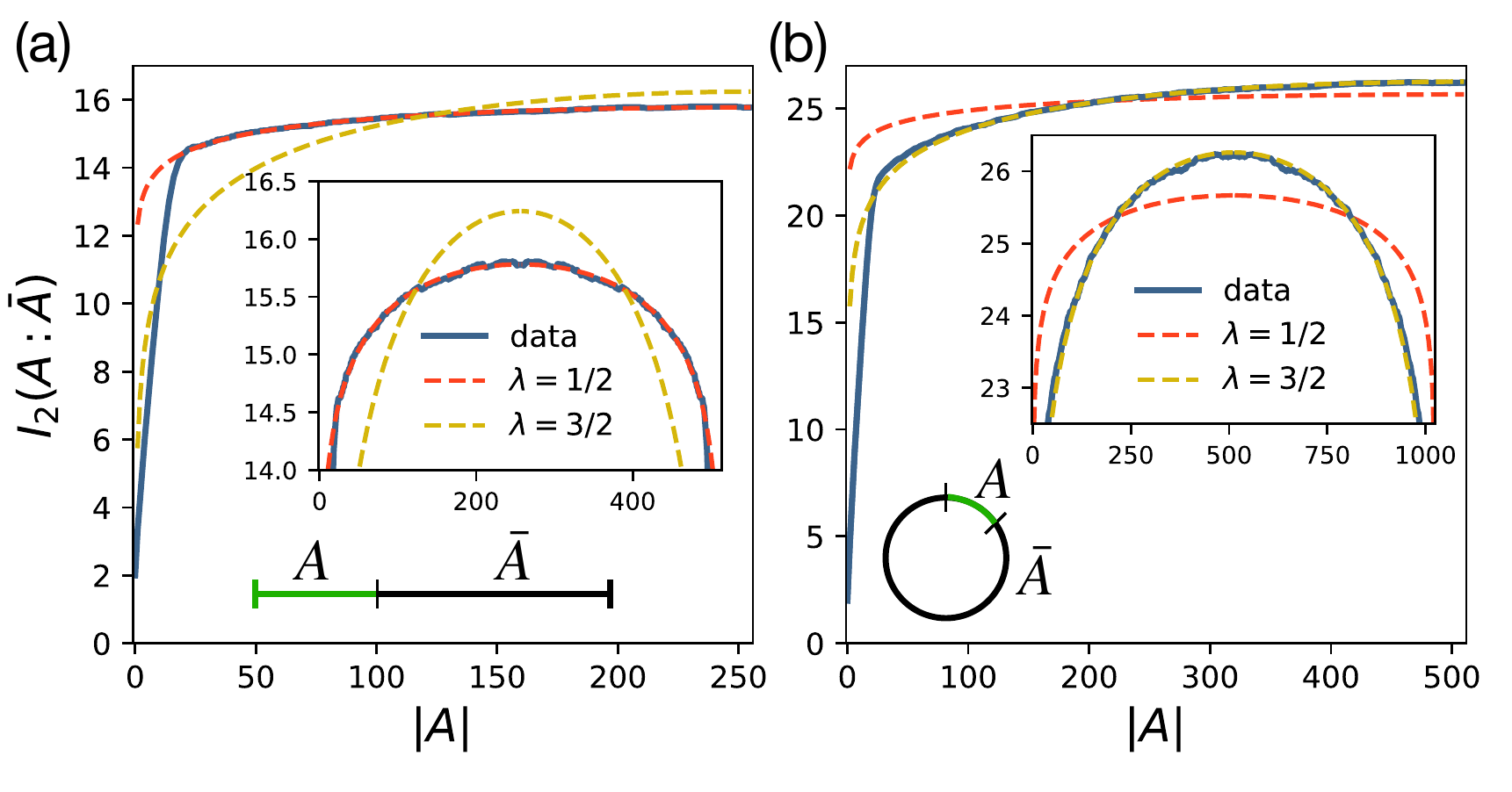}
    \caption{Mixed purification phase in spacetime duals of Haar-random circuits. Stabilizer simulations of $\tilde{L}$ initially-mixed qubits evolving under the spacetime-dual of a random unitary circuit for time $\tilde{T} = 4\tilde{L}$, averaged over $6\times 10^5$ realizations. 
    Annealed average of the Renyi-2 ``mutual information'' $I_2(A:\bar{A}) = S_2 (A)+S_2(\bar{A}) - S_2(A\bar{A})$, where $A$ is a contiguous subregion that is 
    (a) near the edge of a system with open boundary conditions, or 
    (b) in a system with periodic boundary conditions (see sketches at the bottom of the panels).
    In both cases we show fits to the functions $f_\lambda(x) \equiv \text{const.}+\lambda \log_2[x(L-x)]$ with $\lambda = 1/2$ and $3/2$.
    The system sizes are $\tilde{L} = 512$ and $1024$ in (a) and (b), respectively. 
    }
    \label{fig:app_Haar}
\end{figure}

In this Appendix we study spacetime-dual circuits from the point of view of \emph{purification dynamics}~\cite{Gullans2020PRX}: i.e. how a fully-mixed initial state $\rho_\text{in}\propto \eye$ gradually loses entropy under the non-unitary dynamics until eventually it becomes pure, $\rho_\text{out} = |\psi_\text{out}\rangle \langle \psi_\text{out}|$.
The time scale for purification, $t_p$, exhibits a sharp phase structure: it is short ($t_p\leq O(\log L )$) in the \emph{pure phase} and long ($t_p=\exp(L)$) in the \emph{mixed phase};
the phases are separaed by critical points where $t_p = \text{poly}(L)$. 
These phases are closely related to the area-law and volume-law phases in pure-state dynamics, respectively~\cite{Gullans2020PRX}. 
In particular the mixed phase describes the emergence of a quantum error correcting code (QECC) capable of hiding quantum information from local measurements for very long times. 
Formally, this QECC is a mixed state $\rho_\text{out}$ produced by running the non-unitary circuit on $\rho_\text{in}\propto \eye$ for a (polynomially) long time, $L\ll T \ll \exp(L)$ (this separation of scales is only well defined in the thermodynamic limit $L\to\infty$). $\rho_\text{out}$ represents a stochastic superposition of all ``codewords'', i.e. all possible pure states one may obtain as outputs by sending pure-state inputs through the circuit.
Properties of this QECC have received much attention recently~\cite{Fan2020}, with an appealing statistical-mechanical interpretation mapping entropies to domain-wall free energies~\cite{LiFisher2020}.
This line of reasoning yields a conjectured universal behavior for the mutual information of the QECC $\rho_\text{out}$ as $I_2(A:\bar{A}) = \frac{3}{2}\log|A|$ ($A$ and $\bar{A}$ denote a contiguous bipartition of the system, shown in Fig.~\ref{fig:app_Haar}). 
This term also shows up as the leading correction (after the volume-law term) to the entanglement in pure-state dynamics: $S(\ell) = s\ell + \frac{3}{2}\log \ell + \dots$. 
As we have seen, this subleading correction to the entanglement is also realized by the spacetime-duals of chaotic unitary circuits in pure-state dynamics. It is thus interesting to ask whether this is reflected in the mixed-state (purification) dynamics as well.

In the following we address this question numerically.
To do so, we note that our analytical discussion applies to the \emph{annealed} average of the entropy: $S^{a} \equiv -\log \overline{\mathcal{P}}$, where $\mathcal{P} = \Tr(\rho^2)$ is the purity.
In a spatiotemporally random circuit, $\rho$ is a linear function of each two-qubit gate $U$ and its adjoint $U^\dagger$; therefore $\mathcal{P}$ is a quadratic form in $U$, $U^\dagger$ whose average over the unitary group is captured by a 2-design.
For this reason, it is possible to compute $S^{(a)}$ by averaging over the Clifford group (which takes polynomial time through stabilizer simulations~\cite{Aaronson2004}), rather than the entire ${\rm U}(q^2)$ group. 

For the rest of this discussion, we specialize to the case of qubits ($q=2$), and measure entropy in bits. 
Given a collection of entropy samples $\{S_i\}$ from $N$ runs of the stabilizer simulation, we have $S^{(a)} \simeq -\log_2 \frac{1}{N} \sum_i 2^{-S_i}$.
We note that this average may be ill-conditioned if the underlying distribution of samples $P(S_i)$ has fat tails (then exponentially rare low-$S$ samples may dominate the average); however, as we will see, this is not the case for uniformly random circuits.
This is also the numerical method used to obtain the data in Fig.~\ref{fig:Haar_num} in the main text.
We direct the reader to other references for details about stabilizer simulations of unitary circuits~\cite{Aaronson2004, Ippoliti2021PRX}; here we only describe the more unusual aspect: edge decoherence.
In the stabilizer formalism the erasure channel can be implemented in two steps, first by acting with a phase-flip error and then with a bit-flip one. 
To do the phase-flip error, we look for any stabilizers with a Pauli $X$ at the edge site: if there are none, nothing happens; if there is exactly one, it is dropped from the stabilizer list; if there are several, $\{g_1,\dots g_k\}$, they are updated as $g_i' = g_1g_i$ for all $i=2,\dots k$, and $g_1$ is dropped. 
The bit-flip error works the same but with $Z$ instead of $X$. 

Numerical results are shown in Fig.~\ref{fig:app_Haar}.
We find that, as in other examples of measurement-induced purification dynamics, the mutual information\footnote{Strictly speaking, the quantity we compute is a combination of Renyi-2 (rather than Von Neumann) entropies, and is therefore \emph{not} a valid mutual information (e.g. it can be negative).} $I_2(A:\bar{A})$ has an area-law term which grows as $\propto \log(t)$ (eventually saturating to a volume law at exponentially late times, when the output state is fully purified) and a sublinearly divergent piece.
This divergent term agrees very precisely with the prediction $(k/2)\log_2\tilde\ell$, with $k = 1$ if the subsystem $A$ is near the edge of a system with open boundaries (Fig.~\ref{fig:app_Haar}(a)) and $k=3$ if $A$ lies in the bulk (implemented here via periodic boundary conditions, Fig.~\ref{fig:app_Haar}(b)). 
Both cases are shown against fits to the symmetrized functions $f_\lambda(x) \equiv {\rm const.} + \lambda \log_2[x(L-x)]$, with $\lambda = 1/2$ and $3/2$.

\begin{figure}
    \centering
    \includegraphics[width=\columnwidth]{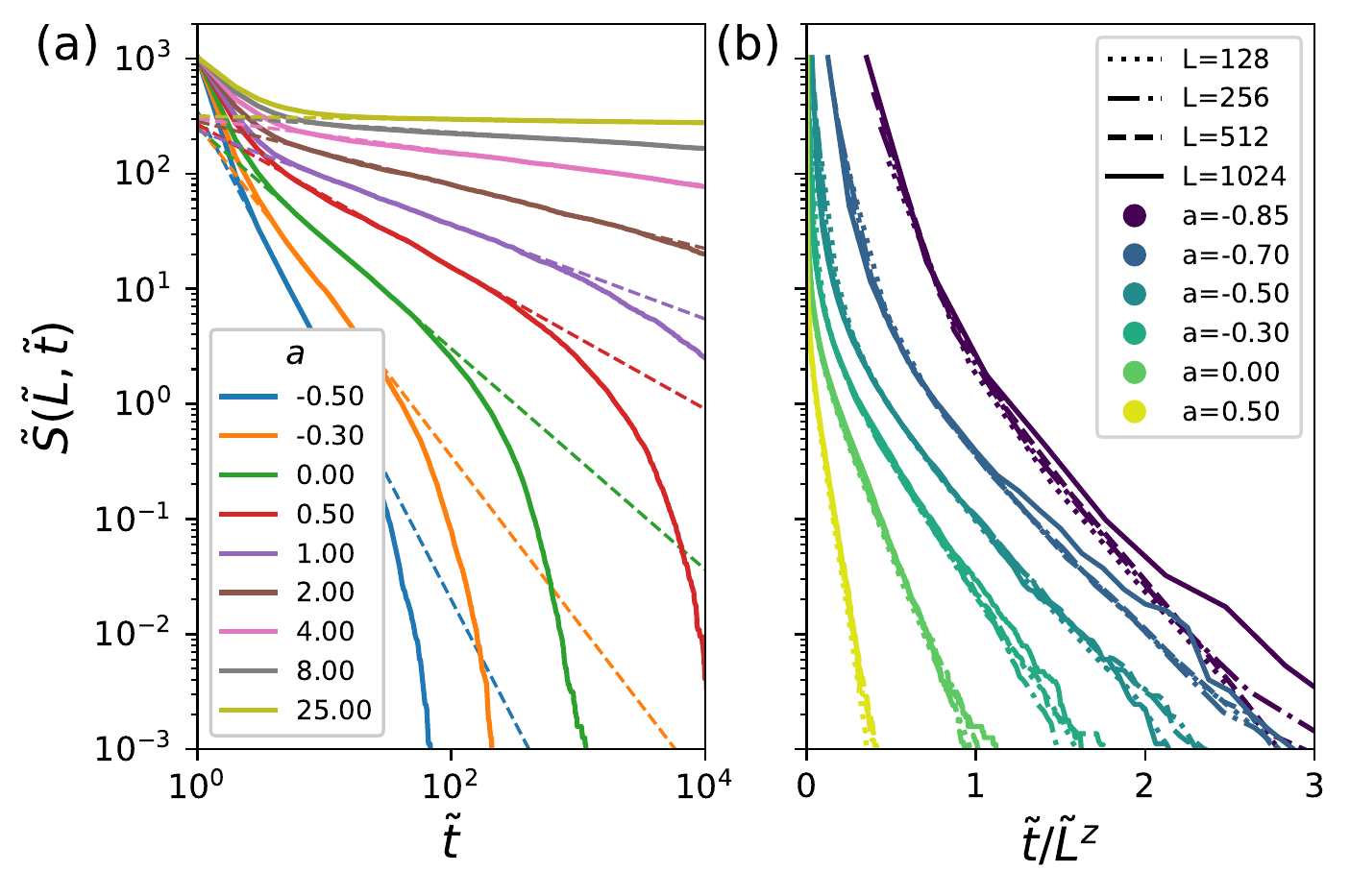}
    \caption{Dynamical purification under the spacetime dual of the Griffiths circuit model with random Clifford gates. 
    (a) Mixed-state entropy $\tilde{S}(\tilde{L},\tilde{t})$ for a system of $\tilde{L} = 1024$ qubits initialized in the fully mixed state, with varying Griffiths parameter $a$ (solid lines). At intermediate times, the entropy drops as a power-law, $\tilde{S}/\tilde{L} \simeq \tilde{t}^{-1/z_p}$ (dashed lines) before crossing over to an exponential form. 
    (b) At each value of $a$, data for different system sizes collapses under $\tilde{t}\mapsto \tilde{t}/\tilde{L}^{z_p}$, with $z_p=a+1$, in both the power-law and exponential decay regimes.
    Numerical data is obtained from stabilizer simulations and averaged over $10^3$-$10^5$ circuit realizations depending on size.}
    \label{fig:griffp}
\end{figure}

Finally, it is interesting to examine the dynamical purification problem in the duals of ``Griffiths circuits''. As we argued earlier, the general understanding of the problem is that the mixed and pure phases should map onto the area-law and volume-law phases in pure-state dynamics; as the fractally-etangled states arising in the duals of ``Griffiths circuits'' fit neither category, the nature of the associated purification dynamics is {\it a priori} unclear.

First, we note that the entropy at time $\tilde t$ can be bounded from above by 
\begin{equation}
\tilde{S}(\tilde{t}) < c \tilde{L} /  \tilde{t}^{1/z_p}\;,
\label{eq:entropy_bound}
\end{equation}
where $z_p = a+1$ and $c$ is a constant. 
This is because after drawing a rate $\gamma$, all but $\gamma \tilde{L}$ of the system's qubits are projectively measured; thus the state's remaining entropy is at most $\gamma \tilde{L}$. 
The intersection of all these bounds for time steps $1,2,\dots \tilde{t}$ yields $\tilde{S}< \tilde{L} \cdot \min_{\tilde \tau = 1,\dots \tilde{t}} \gamma_{\tilde\tau}$; as the minimum $\gamma_{\tilde \tau}$ typically scales as $\tilde{t}^{-1/(a+1)}$, we recover the bound in Eq.~\eqref{eq:entropy_bound} above.

This power-law scaling rules out a mixed phase (where purification is exponentially slow). The question is whether the power-law bound above is saturated or not.
Numerical simulations of Clifford circuits with the stabilizer method answer this question in the affirmative; results are shown in Fig.~\ref{fig:griffp}. 
In particular, we find that the entropy collapses on a scaling function $\tilde{S} \simeq g(\tilde{t} / \tilde{L}^{z_p})$, with $g$ obeying $g(x) \sim x^{-1/z_p}$ at small $x$ (see Fig.~\ref{fig:griffp}(a)), which saturates the bound in Eq.~\eqref{eq:entropy_bound}, before crossing over to an exponential decay $e^{-cx}$ at large $x$ (see Fig.~\ref{fig:griffp}(b)). 
At $z_p = 1$ (achieved here by $a=0$), this behavior ($\sim 1/x$ to $e^{-cx}$) agrees with what is found generically at measurement-induced transitions, namely collapse of the entropy onto a single-parameter dependence on the cross ratio, a signature of CFT behavior~\cite{Li2020}. 
At other values of $a$, this family of models thus appears to realize a range of critical purification dynamics described by anisotropic CFTs with arbitrary $z \equiv a+1 \in (0, \infty)$.


\section{Nature of the non-thermal volume-law entangled phase \label{app:haar}}

Our results in Sec.~\ref{sec:Haar} and Appendix~\ref{app:purif} regarding the non-thermal volume-law entangled phase in spacetime-duals of chaotic unitary circuits raise an interesting question.
Given that the $(3/2)\log\tilde{\ell}$ subleading correction we identified is in agreement with predictions for the analogous phase in general unitary-measurement circuits, it is natural to ask whether the two phases are the same. 
If so, the spacetime-duals of Haar-random circuits would provide an exact microscopic realization of the ``capillary wave theory'' of quantum error correcting codes presented in Ref.~[\onlinecite{LiFisher2020}].

To address this question, we consider a distinct diagnostic introduced in Ref.~[\onlinecite{Fan2020}]: 
the mutual information between a single qubit, located at a variable position $x$ inside a finite subsystem $A$, and the (ideally infinite) complementary subsystem $\bar{A}$, $f(x) \equiv I([x]:\bar{A})$. 
This is a measure of how much entanglement can be destroyed by a single-site measurement inside $A$. Intuitively, this should be a decreasing function of $x$ (taking $x=0$ to be at the edge of $A$). 
For the stability of the volume-law phase in unitary-measurement circuits, $f(x)$ should decay sufficiently fast, so that its integral does not diverge (a necessary condition for measurements to destroy no more than an $O(1)$ amount of entanglement per time step, which can in principle be compensated by the unitary gates acting at the boundary of $A$). 
In generic unitary-measurement circuits this quantity was conjectured to scale as $f(x) \sim x^{-3/2}$~\cite{Fan2020}, but recent work on Clifford circuits\cite{LiVijayFisher2021} finds instead $f(x) \sim x^{-1.25}$.

\begin{figure}
    \centering
    \includegraphics[width=\columnwidth]{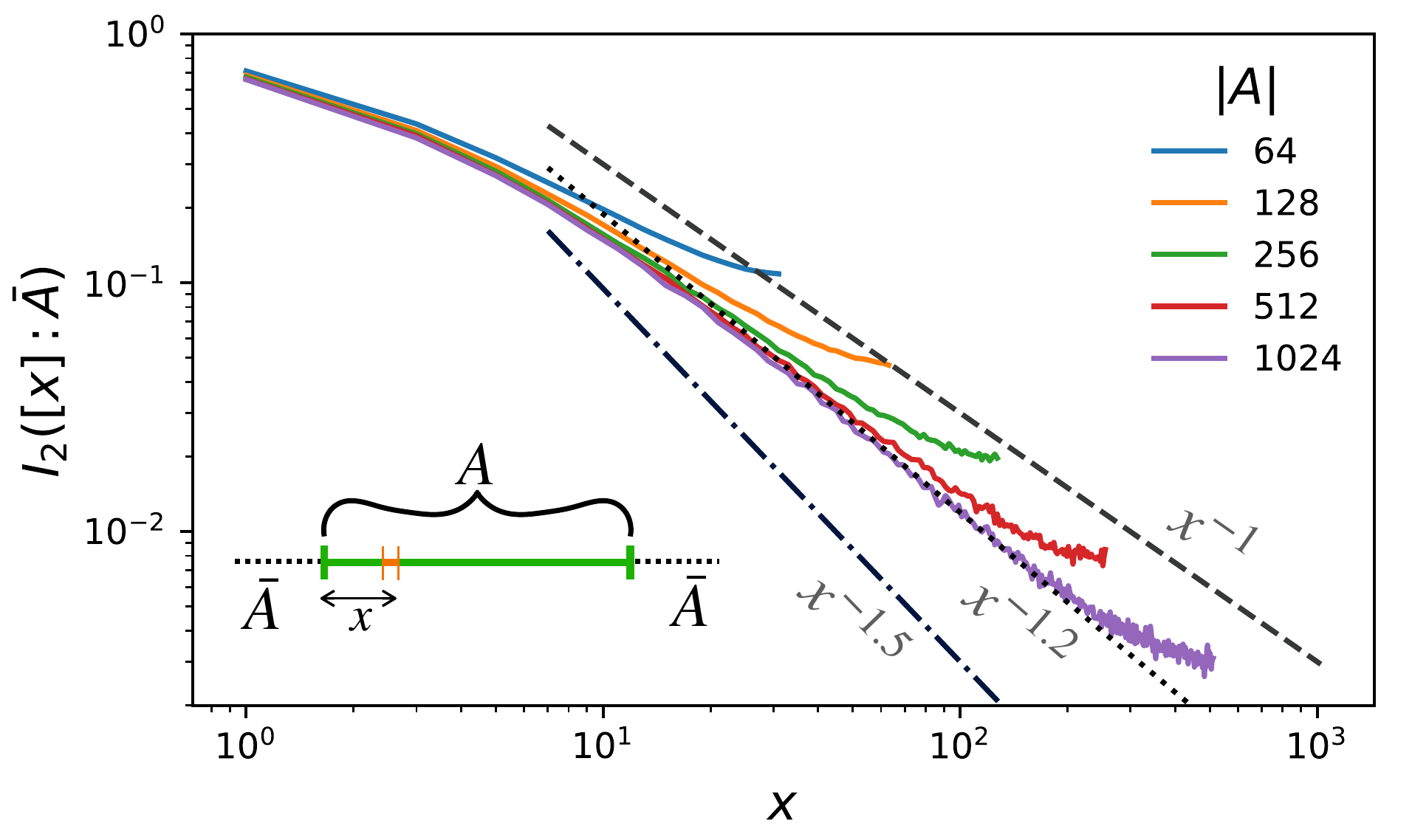}
    \caption{Renyi-2 ``mutual information'' between a single qubit inside a contiguous region $A$ and the (infinite) complement $\bar{A}$, as a function of the distance $x$ between the qubit and the boundary of $A$. 
    Numerical data from stabilizer simulations of the spacetime-dual of a random Clifford circuit (showing only $x<|A|/2$); annealed average over $2\times10^4$-$5\times 10^5$ realizations depending on size. 
    At $|A| = 1024$ the best fit to a power law decay is $\sim x^{-1.2}$ (dotted line), though there is a substantial finite-size drift towards larger exponents. Also shown for comparison are power-laws $x^{-1}$ (dashed line) and $x^{-1.5}$ (dot-dashed line). }    \label{fig:MI}
\end{figure}

To compute this quantity in our setting, we consider a finite subsystem $A$ embedded in an infinite system, initialized in a Bell pair (pure) product state, and evolved under the dual of a random unitary circuit for a long time; 
in the thermodynamic limit, tracing out $\bar{A}$ yields a cancellation similar to the one we used for the pure-state entanglement calculation in Fig.~\ref{fig:boundary_dissip}, which elides all of the circuit except the portion contained inside a ``light cone'' terminating at $A$. 
One is left with a finite-size calculation on subsystem $A$ only, with a fully mixed state and ``light-cone'' boundary conditions as in Ref.~[\onlinecite{Ippoliti2021PRL}]. 
It is then possible to compute $f(x)$ from the resulting mixed state on $A$ by exploiting the fact that $A\cup \bar{A}$ is in a pure state and using
\begin{align}
f(x) 
& = I_2([x]:\bar{A}) = S_{[x]} + S_{\bar A} - S_{[x]\cup \bar{A}} \nonumber \\
& = S_{[x]} + S_{A} - S_{A\setminus [x]}
\end{align}
All three entropies in the expression are computable from $\rho_A$ without reference to $\bar{A}$. 
Hence our calculation implicitly takes the $\bar{A}\to\infty$ limit.

The numerical results for subsystems $A$ of up to 1024 qubits, shown in Fig.~\ref{fig:MI}, suggest that the best fit to a power law $\sim x^{-\lambda}$ is $\lambda \simeq 1.2$.
This is in remarkable agreement with the results of Ref.~[\onlinecite{LiVijayFisher2021}], which finds an exponent $\lambda \simeq  1.25$ in an almost identical quantity, strongly suggesting that the two volume-law phases (in generic monitored circuits and in duals of chaotic circuits) are in fact the same.


\section{Random-walk entanglement computation for Griffiths circuits\label{app:griffiths_rw}}

In this Appendix we present results on entanglement in spacetime-duals of the ``Griffiths circuits'' discussed in Sec.~\ref{sec:fractal} based on an alternative method, which also allows us to analyze more carefully the role of corrections to scaling.

We consider the random-walk computation of the annealed average of the entanglement entropy $S^{(a)}$ for Haar-random circuits, as discussed in Sec.~\ref{sec:Haar}.
The idea there is to take advantage of the edge decoherence picture and adapt a method developed for entanglement growth in unitary circuits, where a recursive formula (Eq.~\eqref{eq:HaarRecursion}) can be used to compute the purity exactly.
In the Griffiths circuit model, as long as the non-identity gates are sampled from the Haar measure, the same approach goes through, up to a small modification to the recursion:
the purity $\mathcal{P}(x,t)$ is either updated as in Eq.~\eqref{eq:HaarRecursion} (if a gate $U\neq \eye$ acts at bond $x$ at time $t$), or left unchanged (if no gate acts). These two occurrences are weighted according to the rate $\gamma_x$ of nontrivial gates acting at bond $x$, giving
\begin{align}
    \mathcal{P}(x,t) 
    & = \gamma_x \frac{q^{-v_E}}{2} [ \mathcal{P}(x-1,t-1) + \mathcal{P}(x+1,t-1)] \nonumber \\
    & \qquad + (1-\gamma_x) \mathcal{P}(x,t-1)
\end{align}
away from the decohering edge, and an analogous modification for the recursion at the edge. 

We remark that evaluating this recursive formula yields the average of $\mathcal{P}$ over both the \emph{location} (whether a gate $U\neq \eye$ is present at bond $x$ at time $t$) and \emph{choice} (which $U \in {\rm U}(q^2)$) of Haar-random gates, given the (quenched) rates $\gamma_x$.
The average $\overline{\mathcal{P}}$ obtained in this way then gives an entropy, $S = -\log_q \overline{\mathcal{P}}$, which can then be averaged over the choice of rates $\{\gamma_x\}$ out of the distribution $P(\gamma) = (a+1)\gamma^a$. 
The resulting average of $S$ is thus partly annealed and partly quenched\footnote{One could also average $\overline{\mathcal P}$ over the choice of rates, getting the overall annealed average of $S$, but this is statistically problematic in the model at hand, where fluctuations of purity over different choices of rates are large.}.

\begin{figure}
    \centering
    \includegraphics[width=\columnwidth]{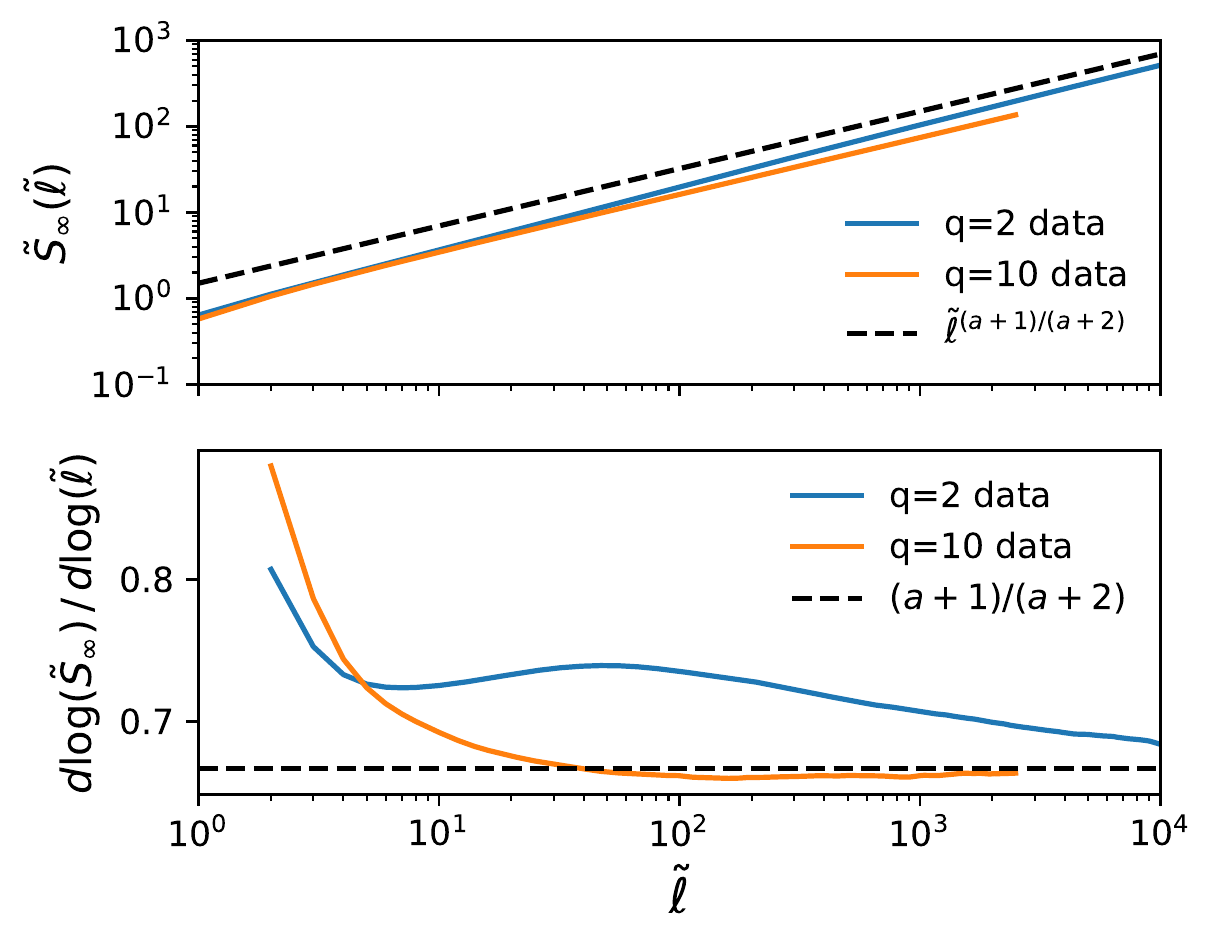}
    \caption{(Top) Entropy $\tilde S$ vs subsystem size $\tilde \ell$ for the Griffiths circuit model with $a=1$, with Haar-random gates, at depth $\tilde{T} = 1024$ (sufficient for saturation to $\tilde{S}_\infty$ at the sizes shown). 
    The data is obtained from the recursive method, for qubits ($q=2$) and 10-dimensional qudits ($q=10$).
    (Bottom) Logarithmic derivative reveals a drift of the data towards the predicted value. The corrections to scaling are large for qubits, but get much smaller at large $q$.}
    \label{fig:griffiths_rw}
\end{figure}

Results obtained by numerically evaluating the recursion formula on $\tilde{T} = 1024$ qubits evolving under edge decoherence are shown in Fig.~\ref{fig:griffiths_rw}, for qubits (the relevant case to compare to the Clifford numerics of Fig.~\ref{fig:fractal}) as well as systems with large local Hilbert spaces ($q=10$), compared to the analytical prediction $\tilde S_\infty \sim \tilde{\ell}^\alpha$, $\alpha = (a+1)/(a+2)$.
A careful look at the size dependence of $\tilde{S}$, e.g. via the logarithmic derivative $d \log(\tilde S)/d\log(\tilde\ell)$, reveals that the asymptotic scaling is consistent with the predicted one, but that corrections to scaling are substantial for qubits, so that the best fit for the exponent in $\tilde S \sim \tilde{\ell}^\alpha$ is significantly off even at large sizes, $\tilde\ell \gtrsim 10^3$. 
(The data shown is for $a=1$, where the predicted scaling is $\alpha=2/3$ and the discrepancy with the finite-size data is worst.)
This suggests that the data in Fig.~\ref{fig:fractal} (from stabilizer simulations) is consistent with the analytical prediction.
A more thorough investigation of subleading corrections, aiming to decouple the effects of (i) unitary dynamics vs edge decoherence, (ii) average over the Haar measure vs the Clifford group, and (iii) quenched vs annealed vs mixed averages, is an interesting question for future studies.
In particular, it would be exciting to detect the logarithmic correction predicted for this model, though the presence of other power-law corrections due to randomness likely makes this task very hard.

\end{document}